\documentclass[12pt]{article}
\usepackage{amsmath,amsfonts}
\usepackage{amscd,amsthm}
\usepackage{geometry}
\usepackage{graphicx}
\usepackage{color}
\usepackage[applemac]{inputenc}
\usepackage{tikz}
\usepackage{graphicx}
\usepackage{color}
\usepackage[applemac]{inputenc}
\usepackage{natbib}
\usepackage{cases}
\usepackage{tikz}
\usepackage{setspace}
\usepackage{hyperref}

\begin{document}

\title{What does the financial market pricing do? \\A simulation analysis with a view to systemic volatility, exuberance and vagary}
\author{Yuri Biondi\thanks{Cnrs - ESCP Europe. Mailing address: 79 avenue de la Republique
75011 Paris. Email: yuri.biondi@gmail.com .} \and Simone Righi\thanks{MTA TK "Lend\"{u}let" Research Center for Educational and Network Studies (RECENS),
Hungarian Academy of Sciences. Mailing address:
Orsz\'{a}gh\'{a}z utca 30, 
1014 Budapest, Hungary.
Email: simone.righi@tk.mta.hu.}}
\date{November 17, 2013}
\maketitle
\begin{abstract}
\cite{biondi2012formation} develop an analytical model to examine the emergent dynamic properties of share market price formation over time, capable to capture important stylized facts. These latter properties prove to be sensitive to regulatory regimes for fundamental information provision, as well as to market confidence conditions among actual and potential investors. Regimes based upon mark-to-market (fair value) measurement of traded security, while generating higher linear correlation between market prices and fundamental signals, also involve higher market instability and volatility. These regimes also incur more relevant episodes of market exuberance and vagary in some regions of the market confidence space, where lower market liquidity further occurs.

\vspace{1cm}



\end{abstract}

\newpage

\setlength{\baselineskip}{20pt}
\singlespacing
\section{Introduction and Related Literature}
\label{intro}
Share Exchanges (financial markets) are socio-economic devices that enable interactions and exchanges between potential buyers and potential sellers while dynamically fixing market prices of exchanged securities. Financial economists have developed elegant models that explain the eventual equilibrium results of this process, without going into the details of price-setting mechanisms. This literature generally identifies two drivers guiding the market price-setting process: matching of supply and demand, and information, with the latter being emphasized by recent contributions on efficient financial markets (\citealt{Bouchaud2009inbook, Shiller2013Sharing, Fama2013Nobel, stiglitz2011heterogeneous,markowitz2005market}). Informationally efficient markets fully and correctly integrate any new (i.e. unexpected) information that affects the fundamental value of traded securities into their price. Due to rational expectations, market prices are expected to change only with unexpected news: informational efficiency implies then that current price ($p_t$) is the best predictor of future ones, making the price time series a random walk. Formally: 
\begin{eqnarray*}
p_{t+1}=E(p_t \vert F_t)\\
F_t=\epsilon_t \text{ with } \epsilon_t \,\,\, \text{i.i.d.} \,\,\, \rightarrow \,\,\, p_t \sim N(\bar{p},\sigma_{\epsilon}).
\label{equilibrium_condition_PROP}
\end{eqnarray*}
Accordingly, fundamental value is supposed to exist and to be correctly known by at least some informed investors who can then act as arbitrageurs between ongoing market prices and that reference value, making a temporary (or illusory) profit while driving back the market price towards its fundamental benchmark, through the very impact of their trades. Indeed the arbitrage mechanism stands at the core of the equilibrium approach to efficient financial markets and can be used to justify the unpredictability of market prices in the long run (\citealt{malkiel2003efficient, lebaron2006agent, iori2012agent, leroy2004rational}). Alternatively, each agent can observe the fundamental value with some noise, which may then be progressively eliminated through emergent information extracted, over time, from evolving market prices and interactions. Following \cite{hayek1945use}, markets can also be studied as aggregators of dispersed information. No investor can then observe fundamental value, but the market aggregating process enables its emergent collective discovery. 

As equilibrium approaches do not disentangle the specifics of the trading process, the interactional nature of its dynamics and its epistemic preconditions, concerns have been raised that this modelization strategy does not correctly lead our understanding of share markets activity especially in the context choosing regulatory designs and regimes (\citealt{Shubik1993INbiondi, Kirman1999InBook, biondi2011problem}). In particular, \cite{sunder1997theory} chapter 7, noted that information from outside the market pricing process needs to be gathered and interpreted by market players in order to be integrated in the market prices; quoting the author: ‘the hypothesis of instantaneous adjustment of price to new information leads paradoxically to the conclusion that such an adjustment cannot occur due to the absence of private incentives to gather information’. Being subject to investigation and interpretation by market players, fundamental value becomes quite an abstract concept which cannot be computed with arbitrary accuracy at every instant in time, with whatever amount of required information available. At the very least, some irreducible margin of judgment and intrinsic error shall remain. As \cite{pareto1983mind} page 30 argued:

\begin{flushright}
In a volume on economics recently published we find that “the price is a concrete manifestation of value.” We are already familiar with the incarnations of Buddha. To them we are now asked to add the incarnations of Value.
\end{flushright}

In contrast, a reasonable (realistic) modelization strategy is to accept that investors idiosyncratically interpret evolving signals of fundamental performance that deliver noisy information about the share-issuing corporate group. As each investor interprets information idiosyncratically, no universal consensus is expected. Investors interact one with another to form their focal price opinions. Trade is then based on disagreement (\citealt{stout2011risk, biondi2011disagreement}) and investors are confronted to the collective dynamic dimension of aggregating supply and demand through the Share Exchange. This dimension introduces an opportunity to profit from the ongoing evolution of aggregate market prices over time. Arbitrage strategies can then be based upon the dynamics of fundamental signals or market prices. A dynamic tension may especially exist between speculative (or chartist) strategies based upon aggregate market prices, and fundamentalist strategies based upon fundamental signals of reference. Furthermore, the process that generates information about the fundamental performance of the share-issuing entity is not a phenomenon neutral to the market dynamics. Fundamental signals are generated through institutional devices external to the market that facilitate its working (\citealt{phelps1987recent, frydman1982towards, fama1992cross}). These institutions provide common knowledge that nurtures fundamental financial analysis processes. Let us define the whole of financial reporting and disclosure, financial analysis techniques and related standards as a common knowledge regime. These regimes assume specific forms, including distinctive accounting regimes that define financial performance and position of the share-issuing corporate group over time. Last but not least, investors factually do not take their decisions in isolation but are subject to social influence and interaction. The study of the impact of this social dimension on agents' behaviour has been developed by a recent strand of economic literature (see \citealt{aldashev2011follies, ozsoylev2011asset, Kukacka20135920}). While this paper does not analyze the dynamic effect of social dimension, it deals with an overall market confidence space that comprises the combination of all the possible states of confidence among and across both sides of the market.

This paper elaborates on the analysis and results by \cite{biondi2012formation} which, drawing upon this modelisation strategy, develop an analytical model where investors’ decision-making is influenced by four dynamic drivers:
\begin{itemize}
\item[(i)] Heterogeneity at individual and group levels, leading to trade on disagreement; 
\item[(ii)] A fundamental information signal generating common knowledge over time; 
\item[(iii)] An emergent social interaction dynamics shaping individual and group beliefs, expectations and forecasts;
\item[(iv)] A market pricing mechanism that aggregates the decisions of investors (orders to sell, to buy or just wait) which disagree according to possibly diverse opinions and profit opportunities. 
\end{itemize}

The present paper assesses the systemic properties of the financial system in terms of market volatility, exuberance, vagary, liquidity and stability, through the analysis of the impact of these drivers on the formation of market prices over time and contexts. These systemic properties prove to be sensitive to common knowledge regimes (which correspond to stylized accounting models of reference for financial reporting and disclosure, namely historical cost and fair value accounting models). Through numerical simulation, we develop a comprehensive economic analysis of the influence of different common knowledge regimes coupled with various combinations of speculative and fundamentalist beliefs across supply and demand sides.
Simulations generate distinctive market price series for every regulatory regime across financial system conditions, enabling comparative analysis of financial systemic performances across time and contexts. 

From a policy perspective, our numerical analysis contributes to the efforts to imagine and design regulatory regimes which show higher degrees of systemic stability and sustainability under a large set of circumstances.

Furthermore, our numerical results provide relevant theoretical points that may further improve conceptual design of control systems such as accounting and prudential regulation. In particular, our disagreement-based analytical model captures and explains \cite{schiller2000irrational}’s market exuberance that is endogenously generated by the market pricing process over space and time. It also provides a theoretical explanation of fair value accounting pro-cyclical contribution to market bubbling (\citealt{enria2004fair, boyer2007assessing}).
Furthermore, these results may be empirically tested against actual share market dynamics, while helping to improve on existing empirical tests for market volatility, exuberance, vagary, liquidity and stability.

The rest of the paper is organized as follows. Section \ref{modelandnotation} introduces the building blocks of the model and related notation. This is followed by a discussion (Section \ref{calibration}) of the calibration used to test the model. The remaining sections provide numerical results regarding the impact of the drivers discussed above on share price formation, pointing to market pricing (Section \ref{PricingDynamics}), volatility (Section \ref{MarketVolatility}), exuberance (Section \ref{exuberance}), and liquidity (Section \ref{liquidity}). Furthermore, information quality is assessed in term of linear correlation with and linear forecasting power of market prices over time (Section (\ref{quality})).

\section{Model and Notation}
\label{modelandnotation}
\cite{biondi2012formation} develop a heterogeneous agents model that generalizes received equilibrium approaches to financial market pricing process. This paper presents results based on an extensive study of a simplified version of that model. This section summarizes it in its main features and in its basic assumptions.

According to \cite{aoki2011reconstructing} chapter 9, two broad categories of chartism and fundamentalism account for most of possible investment strategies. Following \cite{hirota2007price} and \cite{heemeijer2009price}, we consider a large population of heterogeneous investors which form their focal price expectations (upon which they base their trading strategies) according to the following generic function: 
\begin{equation}
E_{i,j,t}(p_{t+1})=p_t+m_{j,t} (p_t-p_{t-1})-\beta_{i,j,t}\delta_{i,j,t}+\gamma_{i,j,t} \phi_i (F_{t}) \,\,\,\, \,\,\,\, \forall i \in [0,1], \forall j\subset (D;S), \forall t
\label{expectation_formation}
\end{equation}
where
\begin{equation}
\delta_{i,j,t} \equiv E_{i,j,t}(p_t)-p_t
\label{correction}
\end{equation}
Each generic investor $i$ can belong to one of two groups $j\subset (D;S)$. Group $D$ is formed by those investors that do not hold shares, implying potential demand, while group $S$ is formed by investors which hold shares (shareholders), implying potential supply.
Equation \ref{expectation_formation} comprises four elements. The first is the past market clearing price $p_t$. 
The second is the signal generated by the market about the aggregate price trend $(p_t-p_{t-1})$. The importance given to this market signal is weighted by the \textit{market confidence} $m_{j,t}$. 
The third element is the individual forecast revision $\beta_{i,j,t}\delta_{i,j,t}$ . It consists of the difference between investor's past price expectation and the last clearing market price that was actually realized (Equation \ref{correction}), weighted by $\beta_{i,j,t}$ which captures both group and individual heterogeneities.
The forth element denotes the formation of an individual opinion based upon available signal of fundamental performance $F_t$, which is common knowledge for both market sides and all the individual investors. This opinion is weighted then by the individual parameter $\phi_i$, while $\gamma_{i,j,t}$ captures both group and individual heterogeneities.

According to this framework of analysis, the financial system is embedded in a dual institutional structure which drives market pricing process: the enterprise entity side is subsumed by the signal of fundamental performance $F_t$, while the market side is captured by the market price trend $(p_t-p_{t-1})$. In particular, the institutional signal that incorporates the fundamental performance (named $Y_t$) of the corporate group whose security is traded is generated by an exogenous mechanism of the type: $F_t\equiv f_t(Y_t)-f_{t-1}(Y_{t-1})$ with $\sum\limits_{h=1}^{t} F_t>0$. This signal is assimilated by each agent idiosyncratically: each investor is then individually characterized by a certain degree of fundamentalism (chartism), captured by the weight $\phi_i$ he attributes to the signal of fundamental performance $F_t : 0 \leq \phi_i \leq 1 \,\, \forall i$. \footnote{Nevertheless, the actual degree of fundamentalism (chartism) for the whole marketplace is endogenously determined by financial system dynamics, and does not depend only on these exogenous subjective attitudes or beliefs.} 

The institutional signal's generating process is inspired by distinctive accounting models of reference, namely Fair Value Accounting (mark-to-market); Historical Cost Accounting and one theoretical benchmark derived from equilibrium approaches, involving target-based signaling provision. The latter regime assumes the very existence and relatively accurate knowledge of fundamental value of reference by informed investors who wish to exploit related signaling. While we may refer to the first two regimes as \textit{actual} regimes, denoting stylized practical modes of accounting, the latter regime can be only thought in a vacuum.

Formally, the \textbf{Historical Random Trend Accounting (HRT)} regime implies an evolving exogenous signal of fundamental performance that is orthogonal to market price dynamics, composed by a stochastic component resulting from positive and negative flows together with a stochastic trend component (\citealt{Biondi2011PureLogic, anthony2004rethinking}), i.e.:\footnote{As robustness check we also simulated our model with a \textit{pure} historical cost approach without the trend: outcomes are extremely similar and relegated to appendix.}
\begin{equation}
F_t=N[-1;+1] + F_{t-1}\cdot U\left[-\frac{1}{2}b,\frac{1}{2}b\right] +  \epsilon_t \,\,\,\, \text{ with } \forall \,\,\,\, b=1;
\label{HRT}
\end{equation}
with $N[-1;+1]$ representing a value extracted from a normal distribution with mean 0 and standard deviation 1, bounded between $-1$ and $+1$ and $U\left[-\frac{1}{2}b,\frac{1}{2}b\right]$ representing a value extracted from a uniform distribution between the two indicated values.

A the opposite side of the spectrum, the \textbf{Fair Value Accounting (FVA)} regime implies a pure mark-to-market accounting system that replicates market information with one time lag (\citealt{kothari2001capital, nissim2008principles}):
\begin{equation}
F_t = (p_{t-1} – p_{t-2}) + \epsilon_t \,\, \forall t.
\label{FVA}	
\end{equation}
Finally, \textbf{Stochastic Target Reverting Accounting (TRA-S)} regime implies a reverting fundamental performance signal that targets a given core value that is bounded within a stochastic band of width $\Delta$ due to white noise error of estimation and other random effects:\footnote{As robustness check we also simulated our model on a signal that reverts toward a precise value: results are very similar and relegated to appendix.}
\begin{equation}
F_t= - (p_t - F_{t-1}) + \Delta + \epsilon_t \,\, \forall t	
\label{TRAS}	
\end{equation}
with $\Delta \sim N(0,1)$ further bounded between $[-1; 1]$. This assumption about $\Delta$ implies that, under this regime, investors know deterministically the range of fundamental values that the traded security can reach.

For all accounting regimes, the same stochastic error is added to account for estimation error, measurement error and other random effects: 
\begin{equation}
\epsilon_t=N[-a;+a] \,\,\,\, \text{with} \,\,\,\, a \geq 0
\label{epsilont}
\end{equation}

The last building block of our model is the mechanism through which the market price is formed at every trade time t. Investors’ bidding strategy is based on their focal price expectations. Actual shareholders can sell or wait for the next period, while potential shareholders can buy or wait for the next period. Accordingly, every shareholder (i.e. each agent $i$ with $\left\{j=S\right\}$) wishes to sell if the clearing market price is higher than his focal price expectations, that is, $p_{t+1} \geq E_{i,D,t}(p_{t+1})$, while every potential buyer ($\left\{j=D\right\}$) wishes to buy if the clearing market price is smaller than his focal price expectations, that is, $p_{t+1} \leq E_{i,D,t}(p_{t+1})$. In particular, extreme investors, which express fully speculative ($\phi=0$) or fully fundamentalist ($\phi=1$) strategies, form their respective focal price expectations as follows:
\begin{equation}
\begin{array}{c}
E_{i=0, j=S} (p_{t+1})=p_t + m_{i=0,S} (p_t - p_{t-1}) + \beta_{i=0,j=S,t} \delta_{i=0, j=S, t} \\
E_{1,S} (p_{t+1})=p_t + m_{1,S} (p_t - p_{t-1}) + \beta_{1,S,t} \delta_{1, S, t} +F_t\\
E_{0,D} (p_{t+1})=p_t + m_{0,D} (p_t - p_{t-1}) + \beta_{0,D,t} \delta_{0,D, t}\\
E_{1,D} (p_{t+1})=p_t + m_{1,D} (p_t - p_{t-1}) + \beta_{1,D,t} \delta_{1,D, t}+F_t
\end{array}
\label{extreme_players}
\end{equation}
At every trading time $t$, the model assumes an aggregate matching process (in line with \citealt{di2012towards, foley1994statistical, anufriev2009asset, chiarella2002simulation, horst2005financial}). The share exchange protocol receives all orders and ranks them on the base of the focal prices expresses by the extreme investors as defined in Equations \ref{extreme_players}. This ranking is based upon the assumption of a uniform distribution of focal prices between the extremes of each market side (formally, $\phi_i \sim U[0,1] \,\, \forall i$). 
This simplifying assumption enables to create a market matching process which generates a complex market dynamics of the market clearing pricing that is analytically treatable (see \cite{biondi2012formation} for further details). Therefore, aggregate demand and supply depend on four focal prices expressed by ideal-type investors over time, defined as follows:
\begin{eqnarray}
\begin{array}{c c}
\overline{P}_t^j=maxarg[E_t(p_{t+1})^j_{i=0}; E_t(p_{t+1})^j_{i=1}] & \forall j \in (D;S) \\
\underline{P}_t^j=minarg[E_t(p_{t+1})^j_{i=0}; E_t(p_{t+1})^j_{i=1}] & \forall j \in (D;S)
\end{array}
\end{eqnarray}
Accordingly, given the imposed uniform distribution of investors over each market side, the Share Exchange Mechanism by aggregating and matching demand and supply delivers the market clearing price, defined as: 
\begin{equation}
p_{t+1}=\begin{cases}p^{NC}=p_t +\eta & \text{if} \,\,\, \overline{P}_{S,t} \leq \underline{P}_{D,t} \\
\frac{\overline{P}_{S,t}(\overline{P}_{D,t}-\underline{P}_{D,t})+\underline{P}_{D,t}(\overline{P}_{S,t}-\underline{P}_{S,t})}{(\overline{P}_{S,t}-\underline{P}_{S,t}) + (\overline{P}_{D,t}-\underline{P}_{D,t})}
& \text{if} \,\,\,\overline{P}_{S,t} \geq \underline{P}_{D,t} 
 \end{cases}
\end{equation}

with $\eta=\frac{N[0,1]}{100}$ being a small tick value that is activated whether demand and offer cannot be matched. In this case a new price is generated that is slightly different from the one of the previous step.\footnote{The presence of this random error turns out to be without material impact on simulation results as discussed in the next section.} Aggregate market price dynamics enriches the passage between the individual and the collective level, making the latter irreducible to the former. Each price pattern becomes then unique over time and space. Replication of several patterns through simulation enables then to infer regularities on the working of this financial system under its distinctive conditions.
 
Finally, regarding our model, one could object to the absence of exchanged quantities (and of individual agents' portfolios). However, individual portfolios are irrelevant here, since we study investors that are not budget constrained and that do form their focal prices on past and next period expectations and posting their orders deterministically by comparing their focal prices with past called price.

\section{Simulation calibration}\label{calibration}
The bulk of our analysis is based on numerical simulations that study the systemic properties of financial market price formation under alternative common knowledge regimes and over the market confidence space. The latter space comprises all the possible combinations of market confidence by investors on both market sides. The analytical model requires calibration to perform this simulation analysis. This calibration does not purport here to obtain realistic assumptions for the parameters, but to improve comparability between various parameter sets and distinctive common knowledge regimes over the overall market confidence space.

In particular, while the measures of market confidence $m_{S,D}$ do change dynamically as result of a social interaction, the present paper neglects this interaction and does analyse simulations in which parameters value do not change over time. This choice is in line with our objective of exploring the systemic effects of different degrees of market confidence.
For the same reason we simplify the general model by imposing the value of $\beta_{i,j,t}=0.5 \,\, \forall i,j,t$ and the value for the variable weighting the individual opinion $\gamma_{i,j,t}=1$. Both calibrations purport to obtain a symmetric setup around the median investor identified as $\phi_i=0.5$. This symmetry choice follows from the hypothesis of uniform distribution between extreme focal prices (on both market sides) and from the fact that all stochastic elements are normally distributed.\footnote{From our assumption that $\phi_i$ is distributed uniformly among the investors between 0 and 1, it follows that values of the market confidence $m_{S,D}>0.5$ imply that the investors tend to be speculative. At the opposite, when $m_{S,D}<0.5$ investors tend to be fundamentalists.} The choice of $\gamma$ further implies that we exclude, for the sake of this analysis, the existence of a systematic, evolutive, bias in the interpretation of $F_t$.

With these simplifications to the model, we run numerical simulations varying $m_S$ and $m_D$ between $0$ and $1$ in steps of $0.01$.
The market price distribution shows a scale-free property. Therefore, for every number of replications, its mean (median) value and variance remains similar for the same market price series length (number of simulated market periods), while, if the number of replications increases, this synthetic measure increases for the same market price series length. Our results are obtained averaging outcomes from series of 500 market prices replicated over 1000 simulation rounds, allowing satisfying comparative analysis between financial systems under distinctive common knowledge regimes. To be sure, if market price distribution is scale-free (power law), these synthetic measures do not characterize the underlying distribution, but, at the same level of series length and replications, they remain comparable between ‘imagined worlds of accounting’ and finance (\citealt{Sunder2011Imagined}).
Given the characteristics of our model, each simulation tells a quite unique ’story’. However, we want to avoid simulation's results depending only on purely transient factors. For this reason, in order to make simulations comparable across types of common knowledge regimes, for each combination of parameters and simulation round, a given random seed is fixed and kept constant whenever the same combination is repeated for another type of common knowledge regime.

In the context of market price formation, whether, at one time step t, demand and supply do not match at all (making thus impossible to generated an updated market price), the auctioneer updates the past market price according to this formula: $P_t=P_{t-1}+0.01 \cdot U(0;1)$. This small error is introduced in order to avoid dead-ends of the market price dynamics. While this device is rarely called under HRT and TRA-S, it activates more often under FVA\footnote{For the FVA this device is called from a minimum of $0.11\%$ of the times up to a maximum of $23,92\%$ of the steps. It is called in average $7.9\%$ of the steps (median $6.76\%$).Detailed about this variable as complementary online material.} especially where the market mood of the two sides is very different. However, where $m_S$ and $m_D$ are not extremely different (that is, the areas where our focus and main findings come from), the auctioneer interventions are very limited.

In order to deal with the possibility that, due to the interaction between the market price dynamics and the fundamental signal dynamics, the market price does fall at (or below) zero, an avoidance mechanism is introduced. Whenever the cumulated fundamental signal (pointing to the underlying fundamental performance) falls at or below zero, this mechanism jump-starts the signal for that period according to the formula: $F_t=F_{t-1} +U[0,1]$. This mechanism is never used except under the FVA, where it is called very rarely (it activates a maximum of $0.02\%$ on the total steps number, with an average activation of $0.00083\%$).

\begin{figure}
\includegraphics[width=0.5\textwidth]{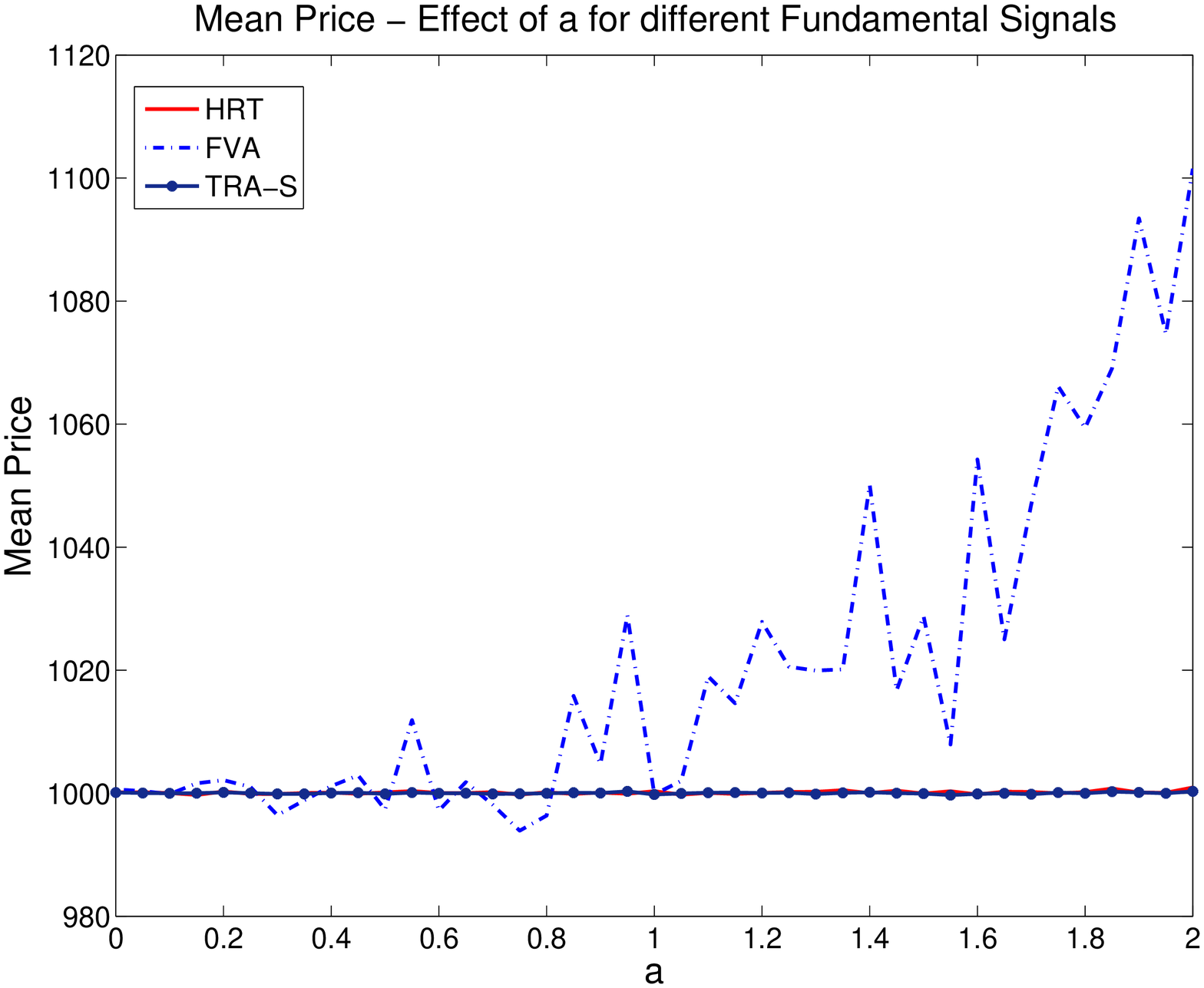}
\includegraphics[width=0.5\textwidth]{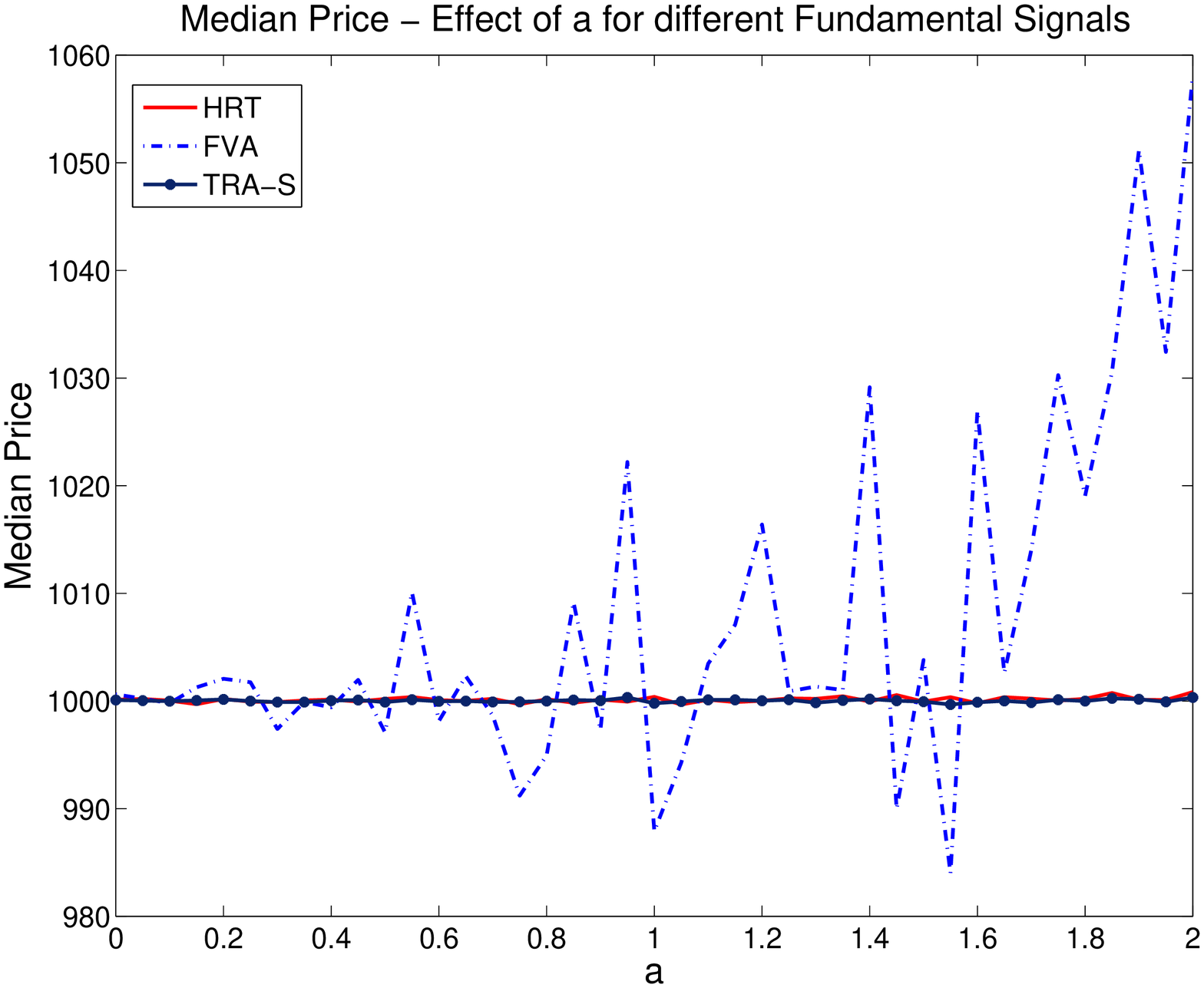}
\label{Impact_of_a}
\caption{Impact of the stochastic error $\epsilon_t$ (from Equation \ref{epsilont}) added to the fundamental signal ($F_t$) on mean (left panel) and median (right panel) price for HRT, FVA and TRA-S common knowledge regimes.}
\end{figure}

As noticed discussing the model, to every fundamental signal, we add an error $\epsilon_t=N[-a;+a]$ with $a \geq 0$. The results of the model are especially sensitive to the magnitude of this error under fair value accounting (FVA), since it is - by assumption - the only exogenous shock that is stochastically added to the financial system under this regime. Figures \ref{Impact_of_a} shows that this error magnifies financial instability already at its first order (mean and median market prices) under this regime. For sake of unbiased comparability of simulation results, we calibrate this error at $a=0.1$, in order to reduce its relative impact for all common knowledge regimes.
Finally, in order to ensure that the results reported below are accurate, we are interested in knowing whether our simulations produce homoskedastic time series. For each simulation, we run the \cite{brown1974robust}'s test for homogeneity of variances (which we chose for his robustness to non-normal data distributions). This test is passed about $95\%$of the times (at a significance level of $5\%$), providing satisfying confidence that our results are robust and that can be tested with regression analyses.

\section{Simulation Results}

\subsection{Market Pricing}
\label{PricingDynamics}
By construction, all common knowledge regimes of fundamental signal provision are designed to swing around the same mean (the initial price is fixed for all simulations to $1000$). Since, all stochastic elements are symmetric around this central point and do not impose any systematic bias to the aggregate market pricing and to investor's beliefs related to the fundamental signal, the mean (median) market price series is then expected to conform to this benchmarking level. Every significative deviation from this central point can only depend on the interaction between the investors’ expectations, the market matching protocol and the evolution of the fundamental signal over time (under each distinctive common knowledge regime). According to Table \ref{1Meanmarketprice_tab}, both HRT and TRA-S produce results that remain around this central point. The FVA signal shows a different result. Observing its distribution of mean and median market price, it emerges that, in some regions of market confidence space, they diverge and reach extreme values.

\begin{table}[h]
\small
\center
\begin{tabular}{| l | c | c | c | c | c | c |}
\hline
Signal Type & Mean $\pm$ Std & Min & Q1 & Median & Q3 & Max\\
\hline
HRT (mean) & 1000.25 $\pm$ 3.6688 & 996.472 & 1000.096 & 1000.237 & 1000.378 & 1035.581\\
HRT (median) & & 996.736 & 1000.098 & 1000.239 & 1000.383 &1016.071\\
\hline
FVA (mean) & 2.4e+60 $\pm$ 2.1e+61 & 999.557 & 1000.490 & 1000.820 & 2.1e+17 & 5.7e+63\\
FVA (median) & & 997.11 & 1000.48 & 1000.816 & 4.2e+08 & 1.2e+32\\
\hline
TRA-S (mean) & 1000.16 $\pm$ 1.7446 & 999.589 & 1000.064 & 1000.154 & 1000.246 & 1000.550 \\
TRA-S (median) & & 999.56 &1000.062 & 1000.154 & 1000.248 &1000.563 \\ 
\hline
\end{tabular}
\caption{Market Prices. Comparison between the distributional characteristics (Mean, Standard Deviation, Minimum value, First Quartile, Median value, Third Quartile and Maximum value) under the different common knowledge regimes. For each regime, the first line indicates the distributional characteristics of the mean value while the second line indicates the distributional characteristics of the median value.}
\label{1Meanmarketprice_tab}
\end{table}

\begin{figure}[h]
\includegraphics[width=0.5\textwidth]{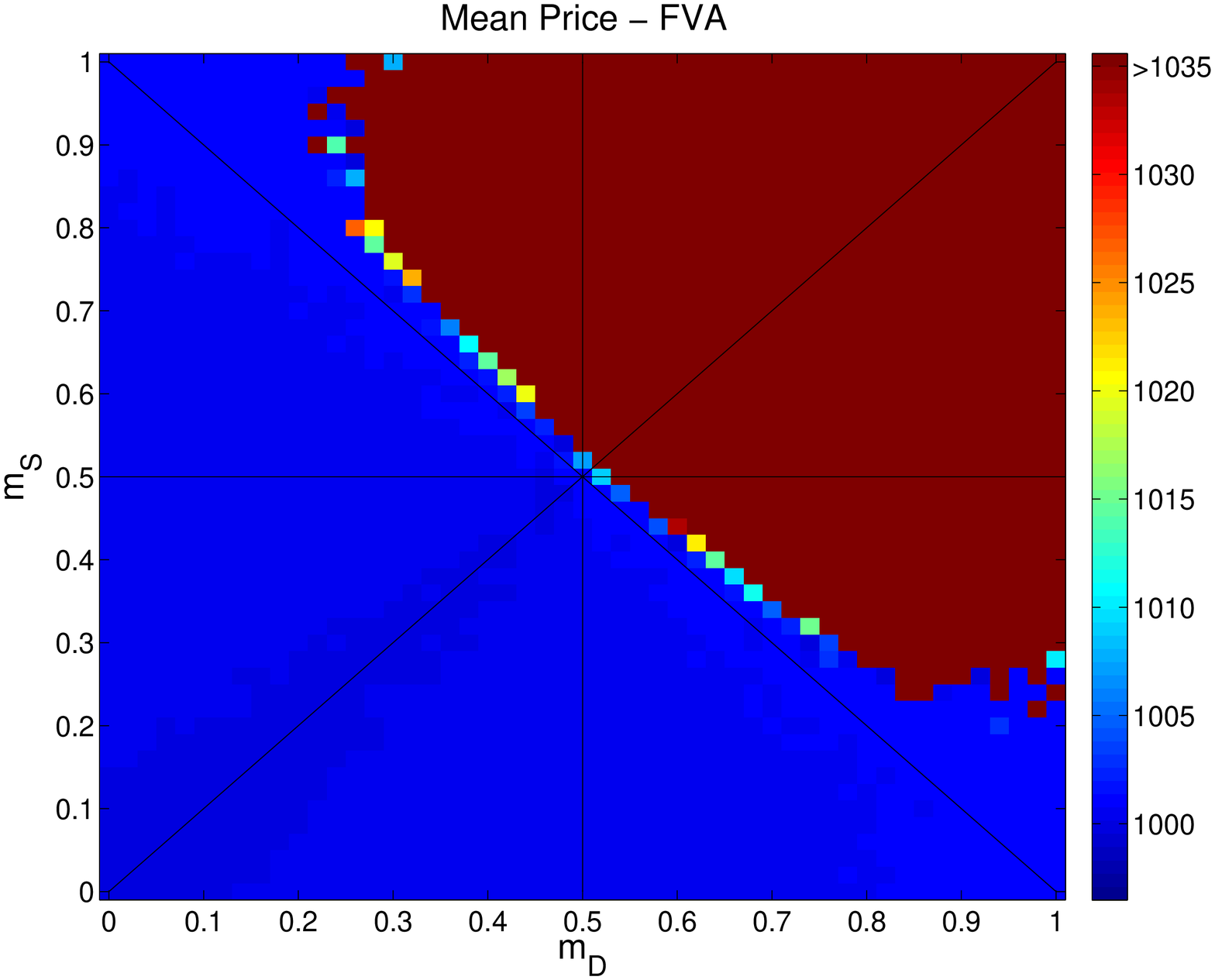}
\includegraphics[width=0.5\textwidth]{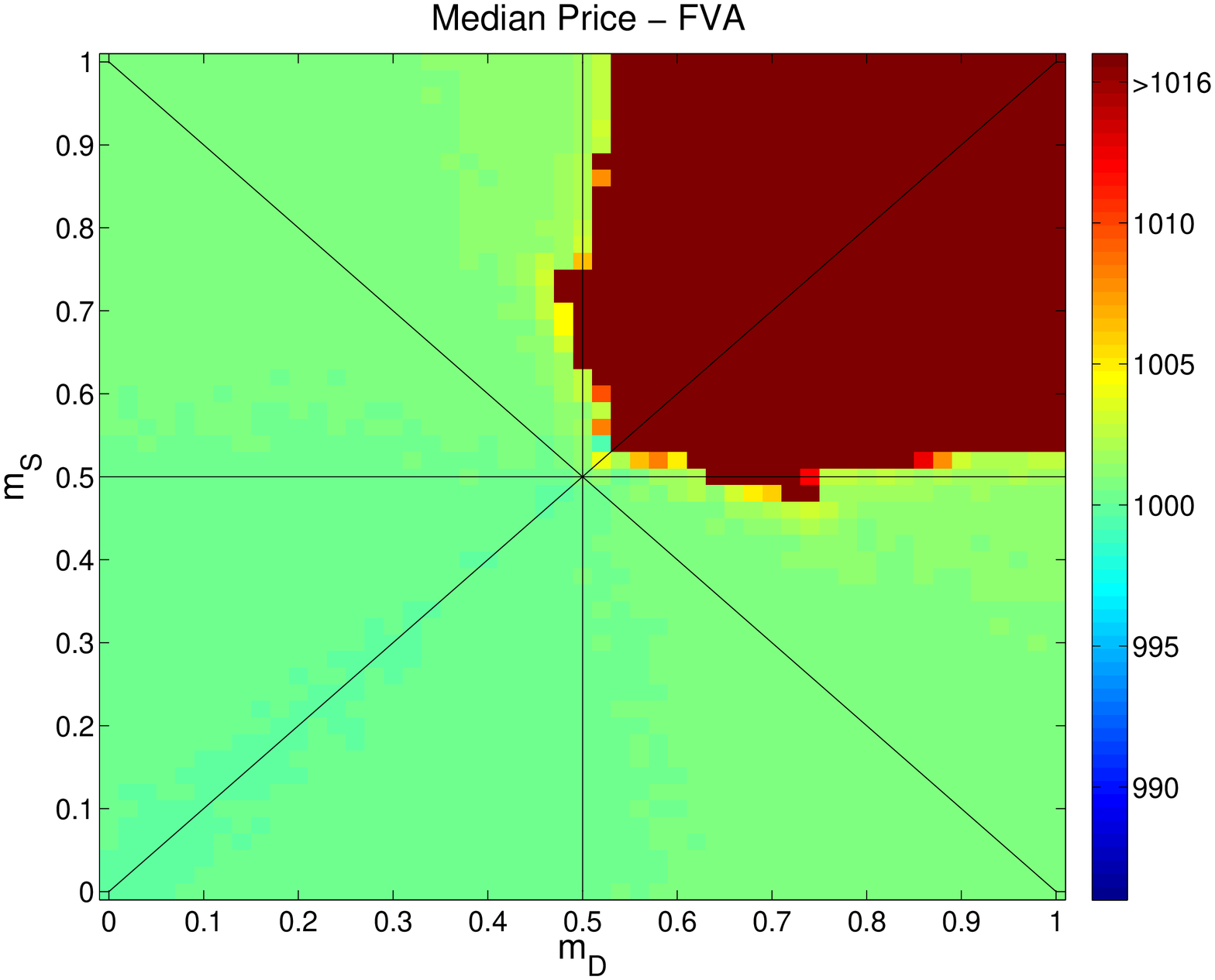}
\caption{Mean Prices (left panel) and Median Prices (right panel) for FVA regime. Dispersion of results over space of market confidence $(m_S; m_D)$. Each data point represents the mean (respectively median) result of 1000 simulations over 500 periods $t$.}
\label{mean_price_FVA}
\end{figure}

In order to gain insights into the reasons for this behavior, it is useful to observe the distribution, over the parameter space, of mean and median market prices in Figure \ref{mean_price_FVA}. Under FVA regime, mean prices significantly diverge from the theoretical central level of 1000 when at least one market sentiment (either the one of supply or the one of demand side) are speculative, implying $m_D \geq 0.5$ or $m_S \geq 0.5$, and especially where both market sides are overconfident in the market signal,  i.e. when $(m_{D} \geq 0.5) \cap (m_{S} \geq 0.5)$. Indeed, where this overconfidence condition is met, bidding by either potential investors (demand) or actual shareholders (supply) drives up the market pricing as long as the market price trend goes up, because the market signal is overweighed - relative to other components - in their forecasting. This creates an overwhelming self-fulfilling and self-reinforcing loop since fundamental signal and market signal reinforce each other, pushing the market price higher and higher. Only very low levels of market confidence by the other side of the market ($m_D$ or $m_S < 0.2$) can counterweight speculative beliefs that drive the market price up. This happens because, where the market sides strongly differ in market confidence, matching between demand and supply orders is reduced (see Figure \ref{Market_ratio_img}). Lower bids on one side of the market tend then to break the loop between market and fundamental signals, keeping the market price dynamics in line with the expected central level. 
Notably, negative bubbling (that is, bubbling dynamics that tends toward zero) is absent in average over our simulations. This happens because our simplified trading strategies do not allow for short selling, thus making it impossible for investors with price decreasing expectations to drive the market price down. Additionally, even when the market trend enters a negative bubble, the downward movement is limited to the minimum bound of zero, while the upward movement is unbounded. Thus, the presence of negative bubbles in the area of overconfidence (found by \citealt{biondi2012formation} studying single simulations) is hidden by the much bigger and unconstrained values generated by positive bubbles.\footnote{The average minimum price, over the FVA regime, is 997.43 (median 999.56) and a minimum of 39.06.}
 
The existence of a  significant distortion of market pricing under fair value accounting regime is comforted by the median of the medians of market price (Figure \ref{mean_price_FVA}, right panel). This measure confirms that the market price distribution is severely strayed away from 1000 under this regime whenever both $m_D>0.5$ and $m_S>0.5$, that is, in the region of joint overconfidence by both demand and supply sides. In the region where only one of them is overconfident, conservative levels by the opposite side do counteract, driving the central level back to the theoretical amount. This provides the additional finding that the formation of market price bubbles in those areas is an extreme event that does not occur at the median level of the distribution.

Overall, this results show that, under some conditions, the FVA is prone to generate relevant market bubbles that materialize in our simulation through persistent explosions of market prices. To be sure, simulated market price levels reach unrealistic values because both investors and the corporate group are not budget-constrained, since a purely theoretical dynamics is under investigation here. These bubbles depend from the self-fulfilling and self-reinforcing interaction between the market sentiment and the fundamental signal which, in the case of FVA, communicates back to the market a fundamental information autocorrelated to the market price dynamics. This result also shows that, not only fair value regime adds instability to the market, but also distorts its working in a self-reinforcing endogenously generated chaos which strays the market pricing away from benchmarking level of fundamental performance. From the regulatory design perspective, this preliminary analysis seems to indicate that introducing FVA may be a harmful choice for financial market stability, as for actual market moods shift endogenously and remain outside control by regulatory authorities. As showed here, shifts toward overconfidence in the market increase the probability of formation of market price bubbles.

\subsection{Market Volatility}
\label{MarketVolatility}
One important characteristic upon which dynamic properties of financial market can be assessed is volatility (pointing to the properties of the market price series at the univariate level). A classic measure of volatility can be computed as the logarithm of standard deviation over the mean of the market price, averaged series by series: 
\begin{equation}
v=log_{10} \left[\frac{\mu_p}{\sigma_p}\right] 
\end{equation}

Table \ref{volatility_price_tab} shows the distributional characteristics of average volatility together with the $75\%$ of the maximum (this latter measure has the objective of obtaining values relative to volatility peak while excluding the most extreme events).

\begin{table}[h]
\small
\center
\begin{tabular}{| l | c | c | c | c | c | c |}
\hline
Signal Type & Mean & Min & Q1 & Median & Q3 & Max \\
\hline
HRT (mean) &0.0036 &0.0017 & 0.0021& 0.0026 & 0.0036 & 0.3363 \\
HRT (75\% Peak) & 0.0045 & 0.0021& 0.003& 0.0032& 0.0045& 0.4790 \\
\hline
FVA (mean) &1.6436 &0.0003& 0.0006 & 0.0009 & 2.9661 & 8.7466 \\
FVA (75\% Peak) & 1.7882 & 0.0004 & 0.0007 & 0.0010 & 4.3298 & 8.7466 \\
\hline
TRA-S (mean) & 0.0017 & 0.0011 & 0.0014 & 0.0016 & 0.0019 & 0.0034 \\
TRA-S (75\% Peak) & 0.0021 & 0.0013 & 0.0017 & 0.0020 & 0.0023& 0.0042 \\
\hline
\end{tabular}
\caption{First Lines: Mean volatility $v_t$ for market price series. Second Lines: Volatility likelihood at $75\%$ of the peak point. Comparison between the distributional characteristics (Mean, Minimum value, First Quartile, Median value, Third Quartile and Maximum value) under the different common knowledge regimes.}
\label{volatility_price_tab}
\end{table}

Even though no actual common knowledge regime can match the performance achieved by theoretical TRA-S, the regime HRT still provides good systemic stability by reducing overall volatility over the whole range of the parameter space of market confidence. This better overall performance is obtained by reducing extreme volatility events, since fair value accounting FVA does have better performances for both minimum volatility, and the first and second quartiles (median). Indeed, the average volatility measures are increased due to the occurrence extreme events with large volatility. Such dramatic events are enabled by FVA but prevented by HRT.
This analysis over the whole distribution of volatility is further corroborated by taking a volatility measure at a certain peak point of this distribution, in line with value-at-risk approaches. Indeed, taking the volatility measure at its third quartile\footnote{This peak point denotes the expected maximum value of volatility under that accounting regime at 75\% likelihood.} (Table \ref{volatility_price_tab}, second lines for each signal) we notice that the FVA shows clearly inferior performance compared to both the theoretical TRA-S and the actual HRT regime.
These results are further corroborated studying the volatility width around the median value, summarized in appendix.
Volatility can be interpreted as a measure of the mis-pricing risk in financial system dynamics. Our results show that, coeteris paribus, FVA increases beyond any fundamental change in the nature of the exchanged security this risk, through the interaction between fundamental signal and market price dynamics.

\subsection{Market Exuberance and Vagary}
\label{exuberance}
Market exuberance points to market prices dynamic that strays away from levels that are reflected in fundamentals. In mathematical terms, this refers to the properties of the market price series relative to the fundamental signal series (at the bivariate level). In order to study this dimension of the financial system, we define the cumulated fundamental signal series $S_t$ that provides an evolving benchmark for ongoing market pricing process over time and space as follows:
\begin{equation}
S_t=\sum_{n=0}^t F_n \forall t \,\,\,\text{ with } \,\,\, F_{n=0} \equiv p_{t=0}=1000
\end{equation}

On this basis, we compute two measures of disconnection between the market price dynamics and this cumulated fundamental signal dynamics:
\begin{itemize}
\item The first measure (Table \ref{exp_max_dist_tab}, first line) is labelled $d_t$ and defined as the relative distance between the current price and its lagged signal of reference, weighted by this latter lagged signal. In order to exclude extreme events from our consideration, we take the third quartile of this measure, providing an assessment of the maximum distance - at 75\% likelihood - generated by the financial system over time. Formally: 
\begin{equation}
Q3[d_t] \,\,\text{with} \,\,d_t \equiv \frac{p_t -S_{t-1}}{S_{t-1}}
\label{Q3dt} 
\end{equation}
\item The second measure, the mean exuberance (Table \ref{exp_max_dist_tab}, second lines), captures an average peak value in market exuberance on the whole simulation sample as follows: 
\begin{equation}
\overline{exub} \equiv \frac{\sum_t (\max d_t - \min d_t)}{T_{max}} \,\, \text{with} \,\, T_{max}=500.
\label{mean_exuberance}
\end{equation}
\end{itemize}
Results regarding these measures are summarized in Table \ref{exp_max_dist_tab}.

\begin{table}[h]
\small
\center
\begin{tabular}{| l | c | c | c | c | c | c |}
\hline
Signal Type & Mean $\pm$ Std & Min & Q1 & Median & Q3 & Max \\
\hline
HRT ($Q3[d_t]$) & 0.0023 $\pm$ 0.0043 &0.0002 & 0.0015& 0.0019 & 0.0023 & 0.2922\\
HRT($\overline{exub}$)&2.10e-05 &2.84e-06 & 1.41e-05& 1.69e-05& 1.98e-05& 1.94e-03\\
\hline
FVA ($Q3[d_t]$) & 0.1018 $\pm$ 0.0160 &0.0001& 0.0003 & 0.0003& 0.1085& 0.8660\\
FVA($\overline{exub}$) &2.22e-04 &9.36e-07 & 2.65e-06& 2.72e-06 & 2.81e-04 & 1.74e-03\\
\hline
TRA-S ($Q3[d_t]$) & 0.0019 $\pm$ 0.0030 &0.0002 & 0.0013& 0.0019& 0.0025& 0.0035\\
TRA-S ($\overline{exub}$)&1.69e-05 &2.79e-06& 1.24e-05& 1.72e-05& 2.12e-05& 2.98e-05\\
\hline
\end{tabular}
\caption{First line: Expected maximum distance $d_t$ at 75\% likelihood (see Equation \ref{Q3dt} for definition). Second Line: Mean exuberance (see Equation \ref{mean_exuberance} for definition). Comparison between the distributional characteristics (Mean, Standard Deviation, Minimum value, First Quartile, Median value, Third Quartile and Maximum value) under the different common knowledge regimes.}
\label{exp_max_dist_tab}
\end{table}

Considering that for complete absence of exuberance this measure would be 0, $Q3[d_t]$ shows that HRT regime remains in line with TRA-S and does not generate any significant amount of exuberance. On the contrary, the FVA shows worse performances, allowing high peaks of volatility in the superior half of the distribution and producing in average a measure of exuberance of $10\%$ around the fundamental signal of reference. Mean exuberance (second lines of Table \ref{exp_max_dist_tab}, for each regime) under FVA is significantly higher than both historical accounting regime and the theoretical one. However, more significative is the fact that the overall performance of FVA is skewed: it performs better than HRT for the inferior half of the sample distribution, but worsens in the superior half, even though the maximum value remains in line with historical cost accounting regimes. This points to the spatial organization of its behavior over the overall market confidence space.
This organization shows that higher exuberance observed under FVA can be attributed to the more likely occurrence of extreme events in which the market dynamics becomes relatively more exuberant. 

Market exuberance is especially important because a persistent distance between the fundamental signal and the market prices over time and contexts implies potential distortion in allocation of resources through the market pricing process. Under exuberant conditions, trades occur either above or below the central point of reference, implying inefficient and unfair transfers of resources across investors and periods.

\subsubsection{Market Vagary (Dynamic Disconnection)}
From a dynamic perspective, market exuberance consists of a overall (comparatively static or cross-sectional) disconnection between market prices and fundamental signals of reference. \citealt{biondi2012formation} find insightful disconnection effects over time between the two dynamics, leading market prices to stray away from their benchmark level of fundamental performance for long durations. It is then interesting to disentangle here these dynamic effects of disconnection conceptually labelled \textit{market vagary} or \textit{errancy} hereafter. For this purpose let define a dissociation measure as: $D_t\equiv p_t-S_t \,\, \forall t$ . For sake of this analysis, we choose a very stringent definition of \textit{dissociation duration}: we count one dissociation duration whenever $D_t$ strays away from its overall mean for more than two standard deviations for at least 10 time periods $t$. We then compute the percentage of time\footnote{Given this strict definition of dissociation we find relatively low percentages of time that fall into this category. This difference, however, preserves significant differences across common knowledge regimes.} periods over the overall length of every simulation in which the fundamental signal is disconnected from the market price (Table \ref{dissperc}, first lines for each regime), as well as the average length of the disconnection durations (Table \ref{dissperc}, second lines for each regime).

\begin{table}[h]
\small
\center
\begin{tabular}{| l | c | c | c | c | c | c |}
\hline
Signal Type & Mean & Min & Q1 & Median & Q3 & Max \\
\hline
HRT ($\%$) &0.3022 &0& 0.2606& 0.3186 & 0.3602 & 0.5078\\
HRT (Periods) &16 &0& 15& 16& 17& 21\\
\hline
FVA($\%$) & 0.2722 & 0 & 0.0456 & 0.3608 & 0.39925 & 0.518\\ 
FVA (Periods) &14 &0& 16& 17& 18& 36\\
\hline
TRA-S ($\%$) & 0.2419 & 0 & 0.1660 & 0.2606 & 0.3246 & 0.4446\\
TRA-S (Periods) &16 &0 & 15& 16& 16& 22\\
\hline
\end{tabular}
\caption{First Lines: Percentage of time in which the market price evolution is dissociated from the cumulated fundamental signal $S_t$. Second Lines: Mean Length of the dissociation periods. Comparison between the distributional characteristics (Mean, Minimum value, First Quartile, Median value, Third Quartile and Maximum value) under the different common knowledge regimes.}\label{dissperc}
\end{table}

\begin{figure}[h]
\center
\includegraphics[width=0.32\textwidth]{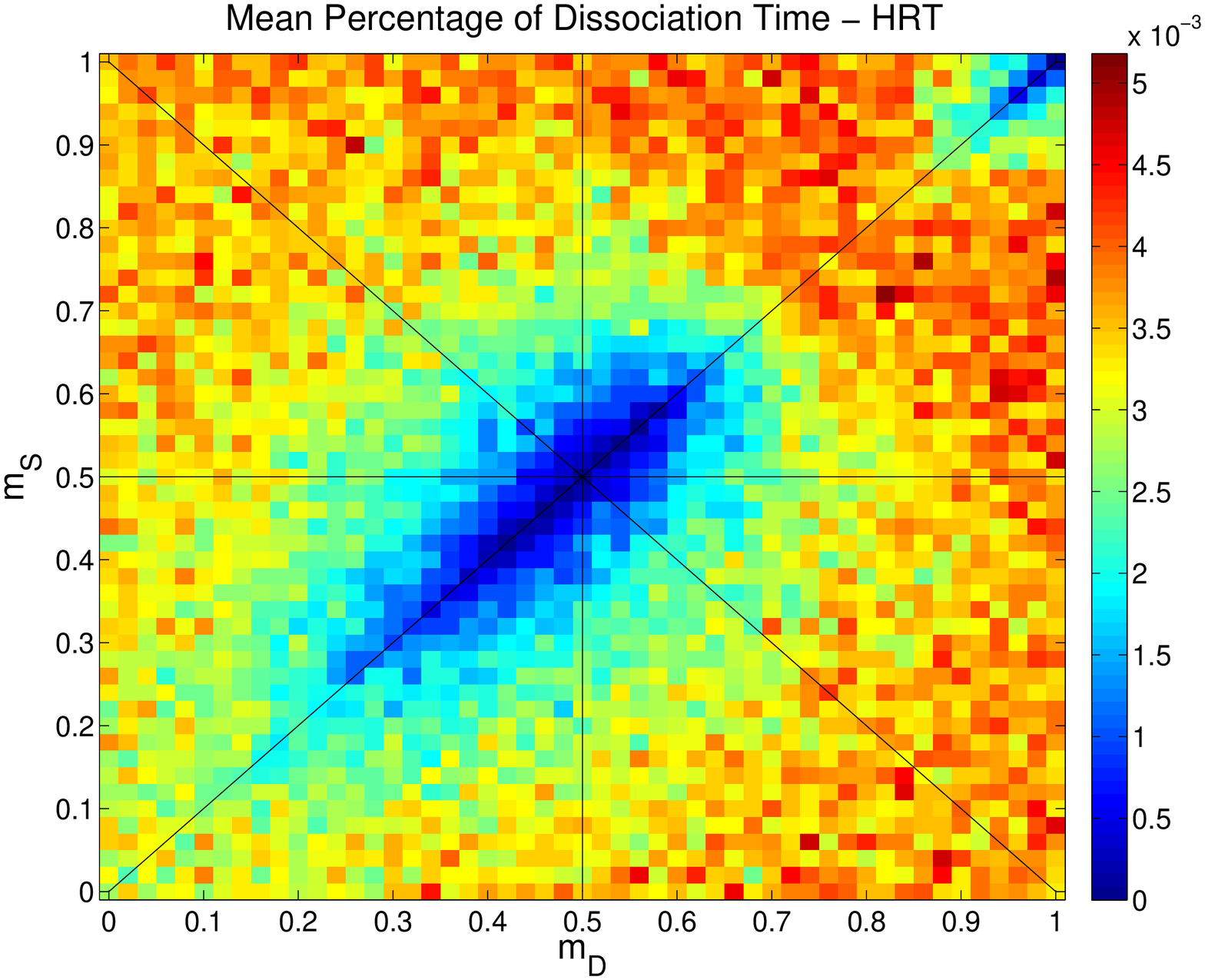}
\includegraphics[width=0.32\textwidth]{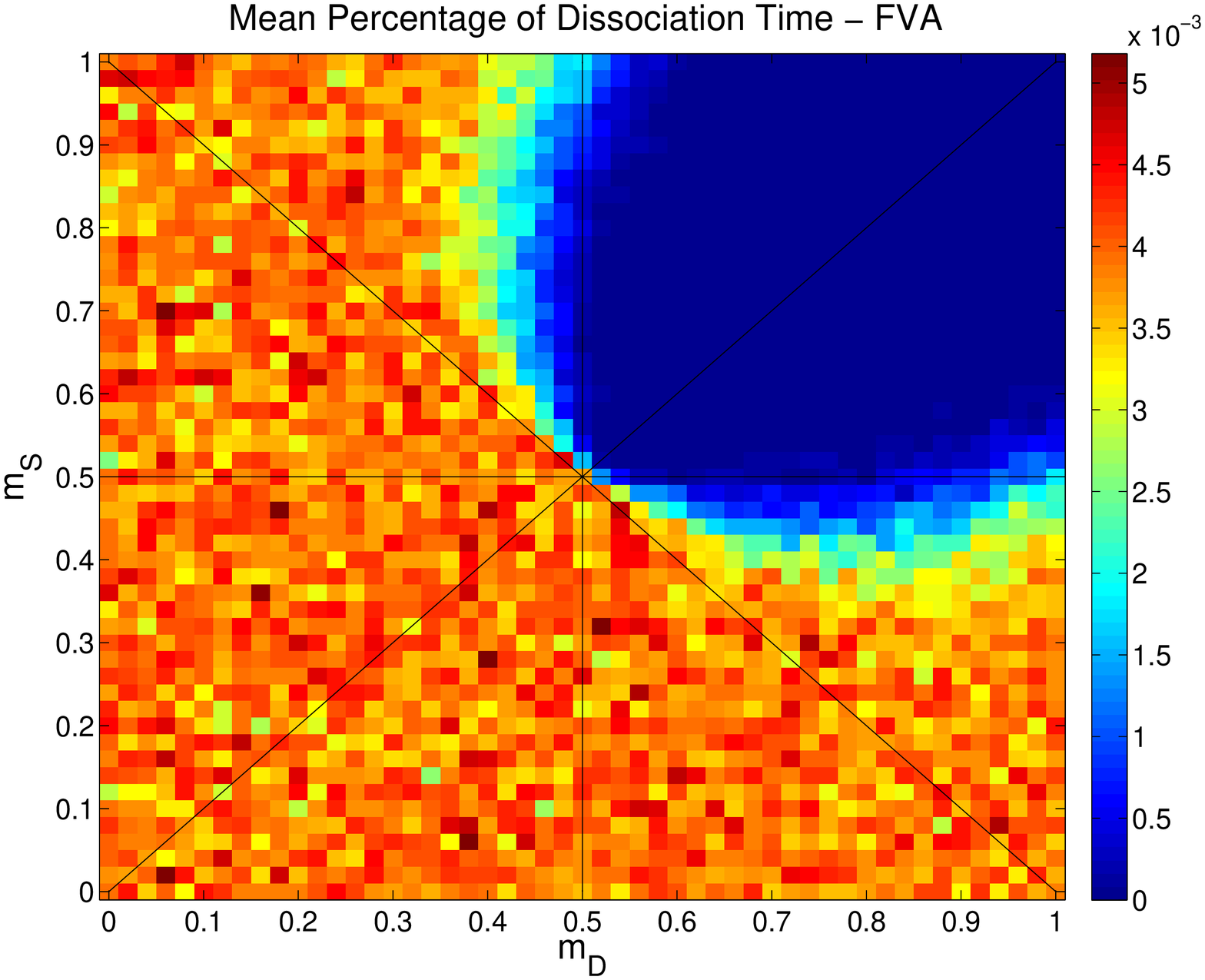}
\includegraphics[width=0.32\textwidth]{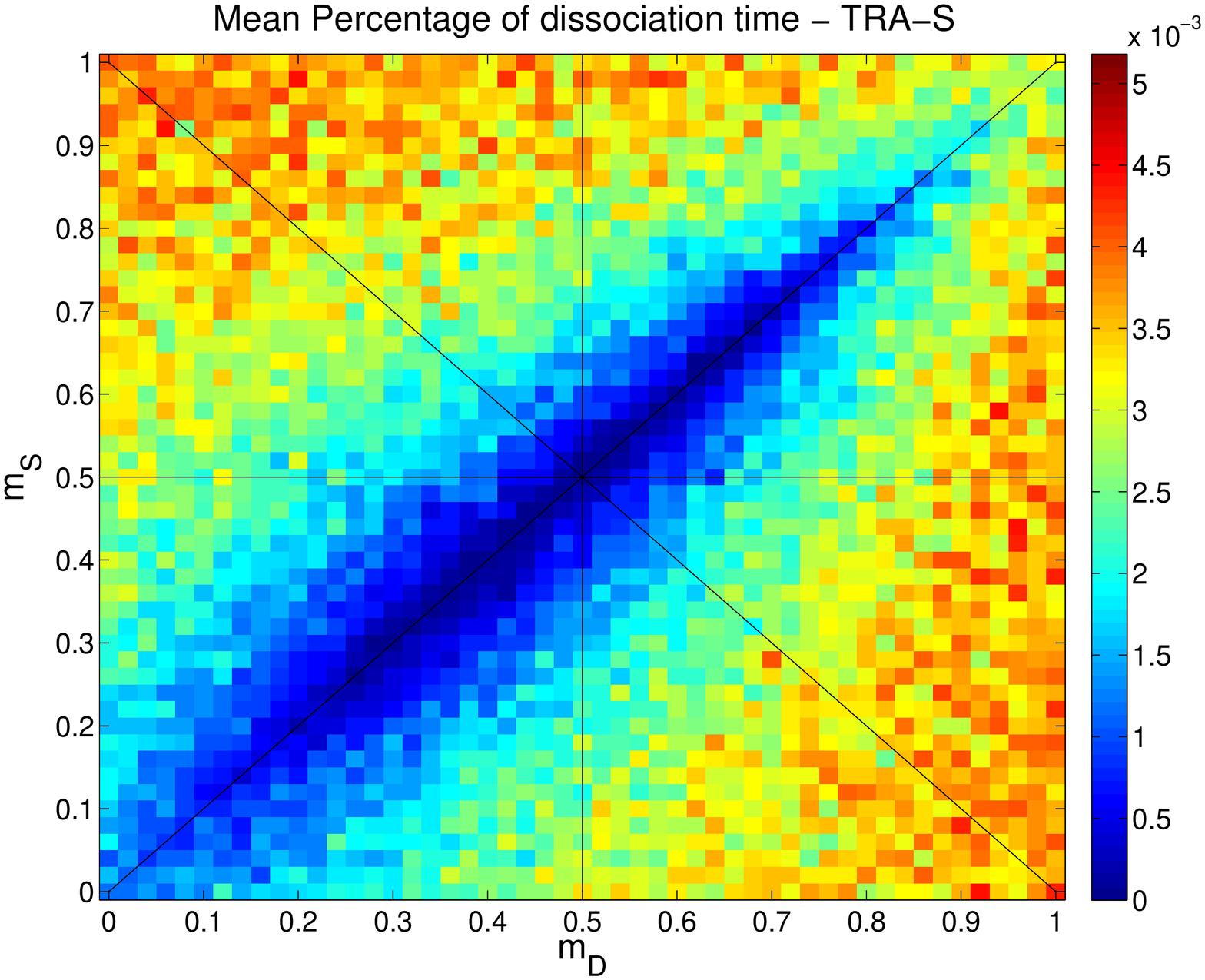}
\caption{Percentage of time in which the market price $p_t$ is dissociated from the cumulated fundamental signal $S_t$ generated by HRT (Left Panel), FVA (Central Panel) and TRA-S (Right Panel). Dispersion of results in space of parameters $(m_S; m_D)$. Each data point represents the average result of 1000 simulations over 500 periods $t$.}
\label{diss_perc_img}
\end{figure}

The dissemination of both measures on the parameter space of market confidence is significant (Figure \ref{diss_perc_img}). Theoretical common knowledge regime TRA-S shows better performances around the parity line where $(m_{j=D,t} \simeq 0.5) \cap (m_{j=S,t}\simeq 0.5) \, \forall \, t$. This spatial organization is crudely replicated by the HRT regime. Under the latter, the percentage of dissociation time remains quite in line with the theoretical benchmark provided by TRA-S. Lower levels are concentrated along the diagonal where market side opinions are quite similar or the same ($m_D \simeq m_S$), especially where both opinions are not too conservative (i.e., outside the corner area where $m_D \simeq m_S < 0.2$). Notably, a specific region of dissociation occurs in the second quadrant, including along the diagonal. In this region, overconfidence in the market signal by both demand and supply is in perpetual tension with fundamental signal HRT that remains independent from market price trend. Speculative bubbles then continuously attempt to occur but they eventually burst because of this dynamic tension. At the opposite, for the corner area (where $m_D \simeq m_S>0.9$), high overconfidence seems to predominate, reducing then the impact of this tension. This phenomenon does not occur under TRA-S, since investors that care about fundamental signals are assumed to treat a common target fundamental value of reference, under this regime. This theoretical regime does not involve stochastic trends in fundamentals that can trigger temporary speculative bubbles over time.

However, it is fair value common knowledge regime FVA that shows the worst performances (higher disconnection durations). This happens everywhere but in the region where volatility is the highest and market prices are distorted at their first order (see Sections \ref{PricingDynamics} on market pricing and \ref{MarketVolatility} on market volatility). In this region, overconfident opinions by both sides of the market continuously agree on upward market price trends, making the market less vagarious but dysfunctional. The market pricing mechanism fixes a clearing price that, even though it remains in line with the fundamental signal, is distorted (since in this area the market price and the signal are auto-correlated by construction). The fundamental signal is then unable to provide any rebalancing benchmark. It cannot act as an ‘accounting lighthouse’ for investors (\citealt{biondi2011disagreement, Biondi2011PureLogic}), leaving them alone with their consensual but distorting beliefs. In this sense, the better performance of FVA is achieved at the cost of the possibility of much higher volatility, exuberance and more likely occurrence of extreme events regarding market price dynamics. In sum, FVA shows its better performance where it becomes unable to drive financial system dynamics in line with benchmark level of fundamental performance, while it shows worst performance elsewhere.

\begin{figure}[h]
\center
\includegraphics[width=0.32\textwidth]{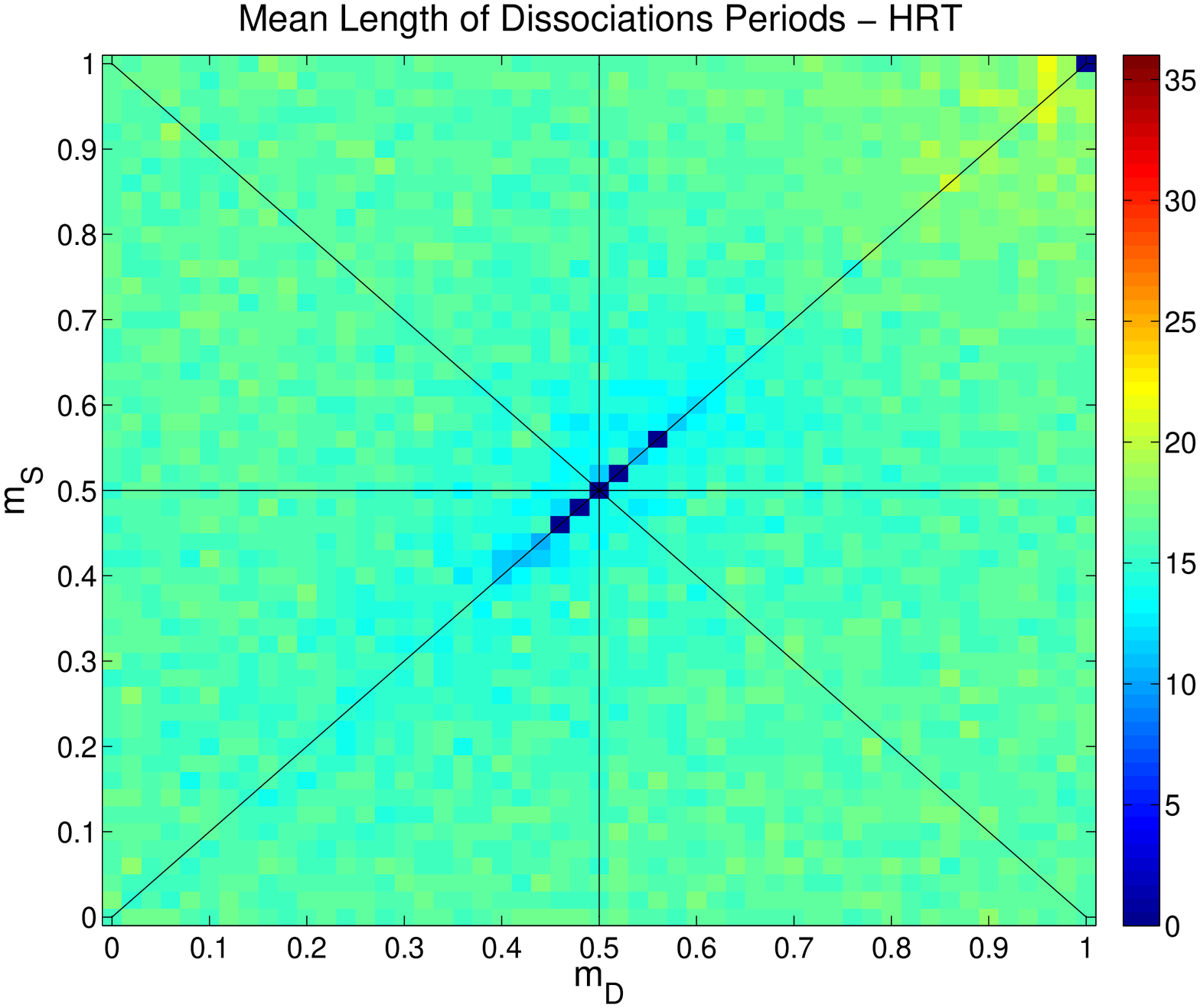}
\includegraphics[width=0.32\textwidth]{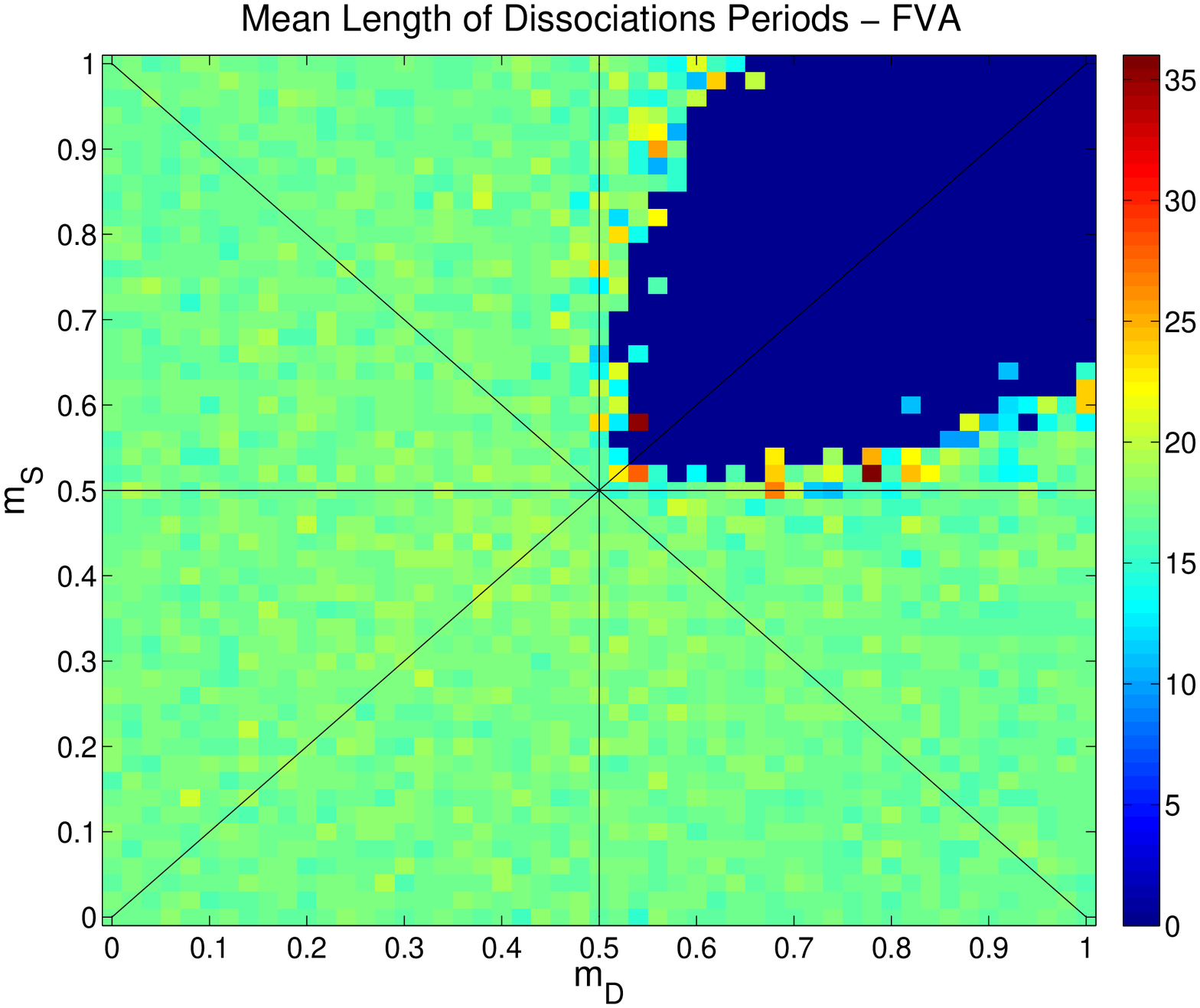}
\includegraphics[width=0.32\textwidth]{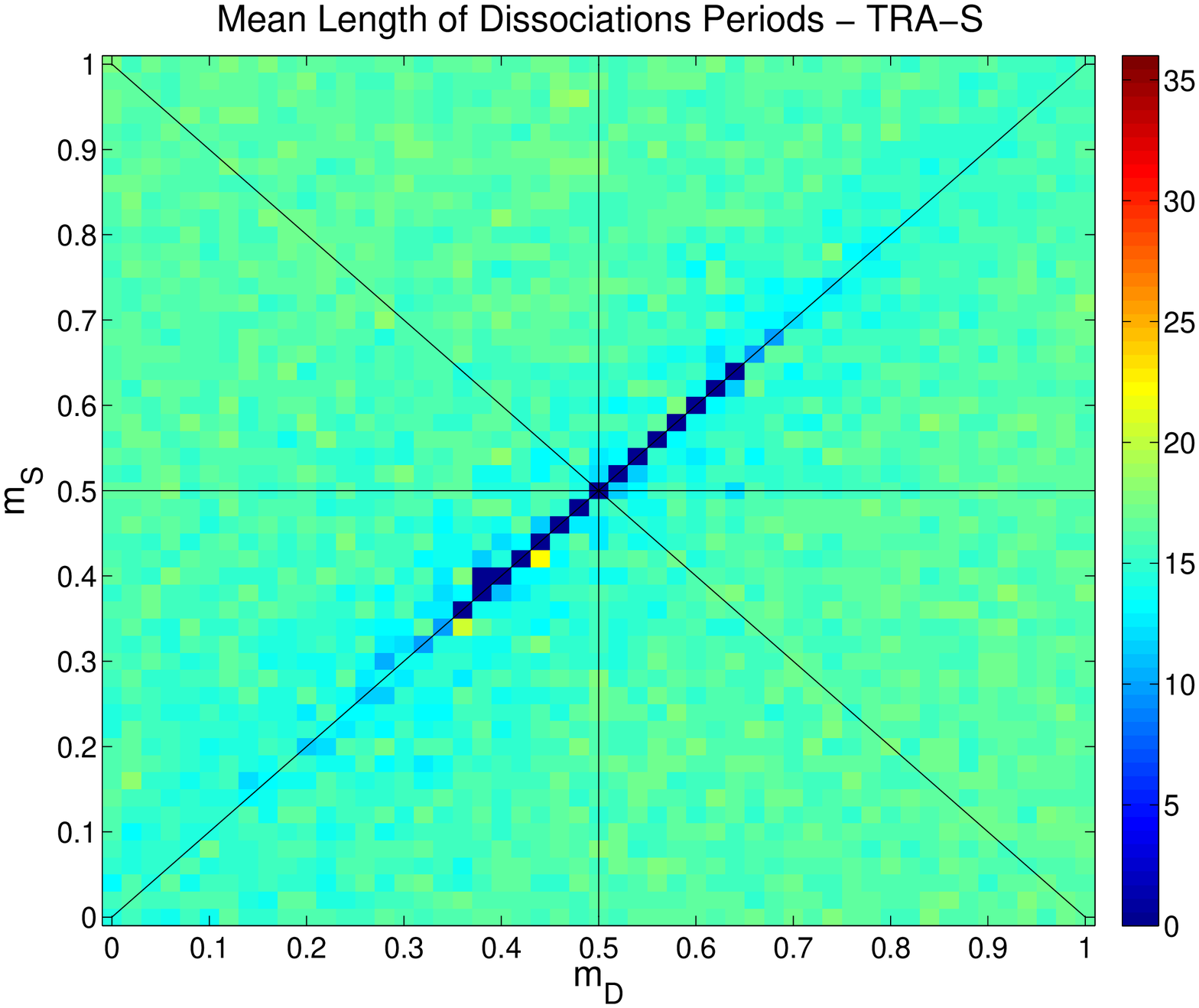}
\caption{Mean Length of dissociation periods for HRT (Left panel), FVA (Central Panel) and TRA-S (Right Panel). Dispersion of results in space of market confidence $(m_S; m_D)$. Each data point represents the average result of 1000 simulations of 500 periods $t$.}
\label{dissoc_length_img}
\end{figure}

This inference is comforted by exploring the mean time length of dissociation (Table \ref{dissperc}, second lines, and Figure \ref{dissoc_length_img}). Again, the performance of FVA in general is obtained by compensating worst results away with those, distorted, from the overconfidence area (second quadrant) of the parameter space of market confidence. On the contrary, all the other common knowledge regimes obtain their best performances (implying shorter dissociation time length) in association with both the parity line and the absolute parity point where $m_{D}= 0.5 \cap m_{S}= 0.5$. 

It is important to study measures of market vagary together with measures of market exuberance since both provide complementary information about the relationship between market price and fundamental signal, clearly showing tendency by FVA to generate market bubbles over that fundamental level of reference. This implies that, not only FVA involves more volatile market price dynamics (as discussed in Section \ref{PricingDynamics}), but may also precludes the market pricing process to properly and timely incorporate fundamental signals. Financial market becomes then \textit{vagarious}, evolving in a fundamentally erratic way that goes even beyond market \textit{exuberance} over those same fundamental levels of performance. 

\subsection{Market Liquidity}
\label{liquidity}

By market liquidity, we mean the ability by the Share Exchange to satisfy trade orders posted by potential investors wishing to buy (potential demand) or sell (potential supply), at any trade time $t$. The degree of market liquidity measures then whether and how much the market matching process enables investors which want to trade to do so.

\begin{figure}[h!]
\center
\includegraphics[width=0.49\textwidth]{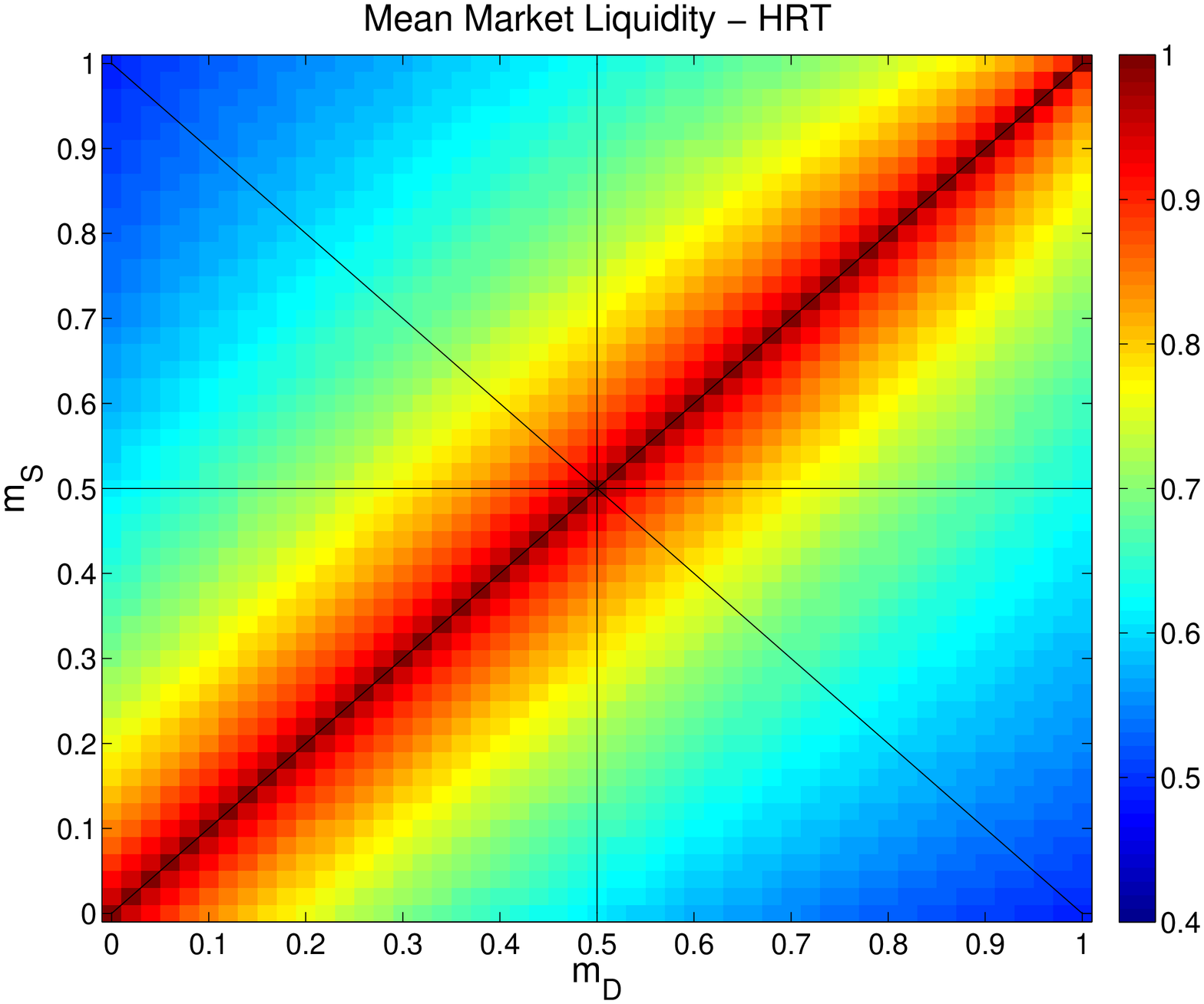}
\includegraphics[width=0.49\textwidth]{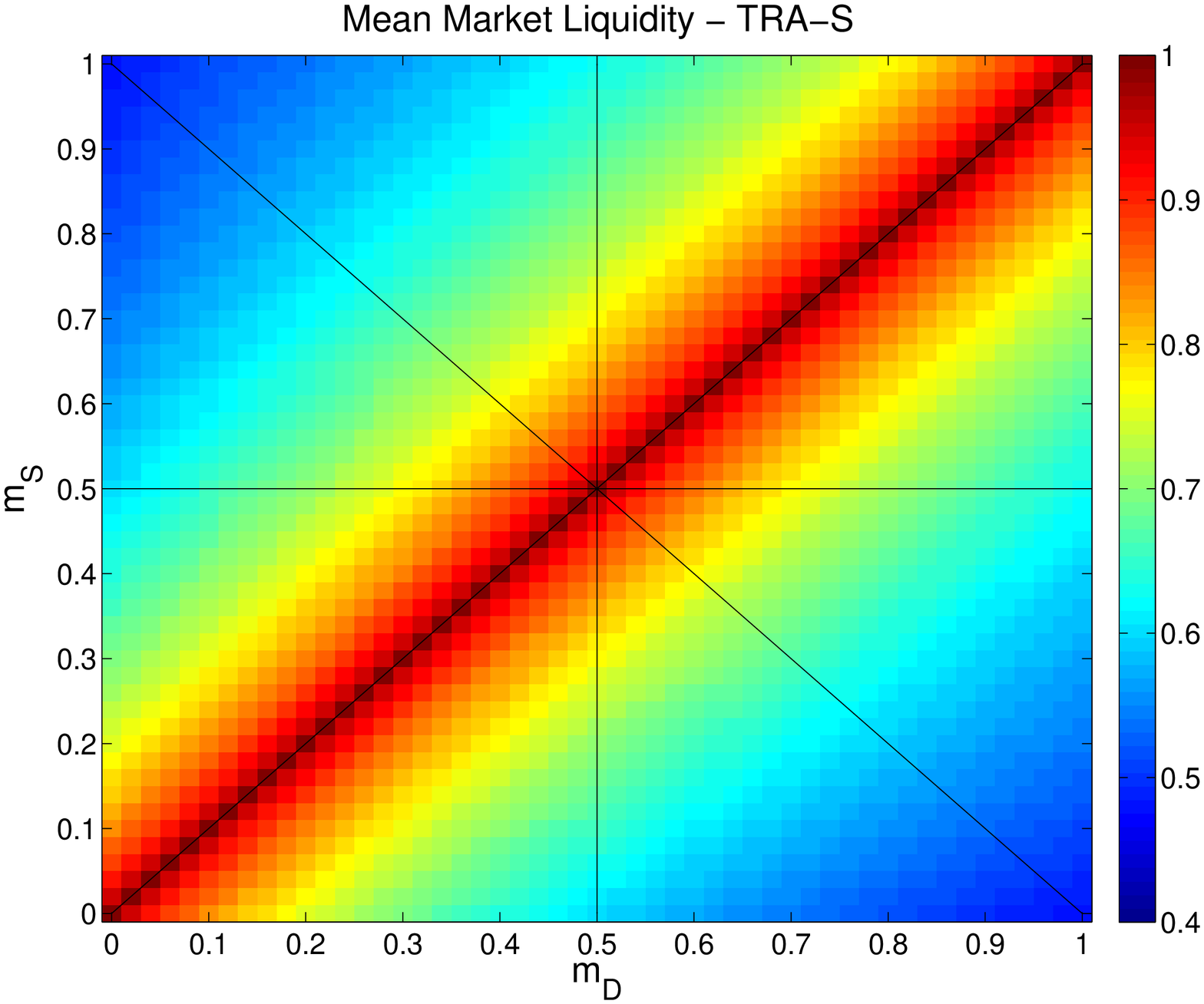}\\
\includegraphics[width=0.49\textwidth]{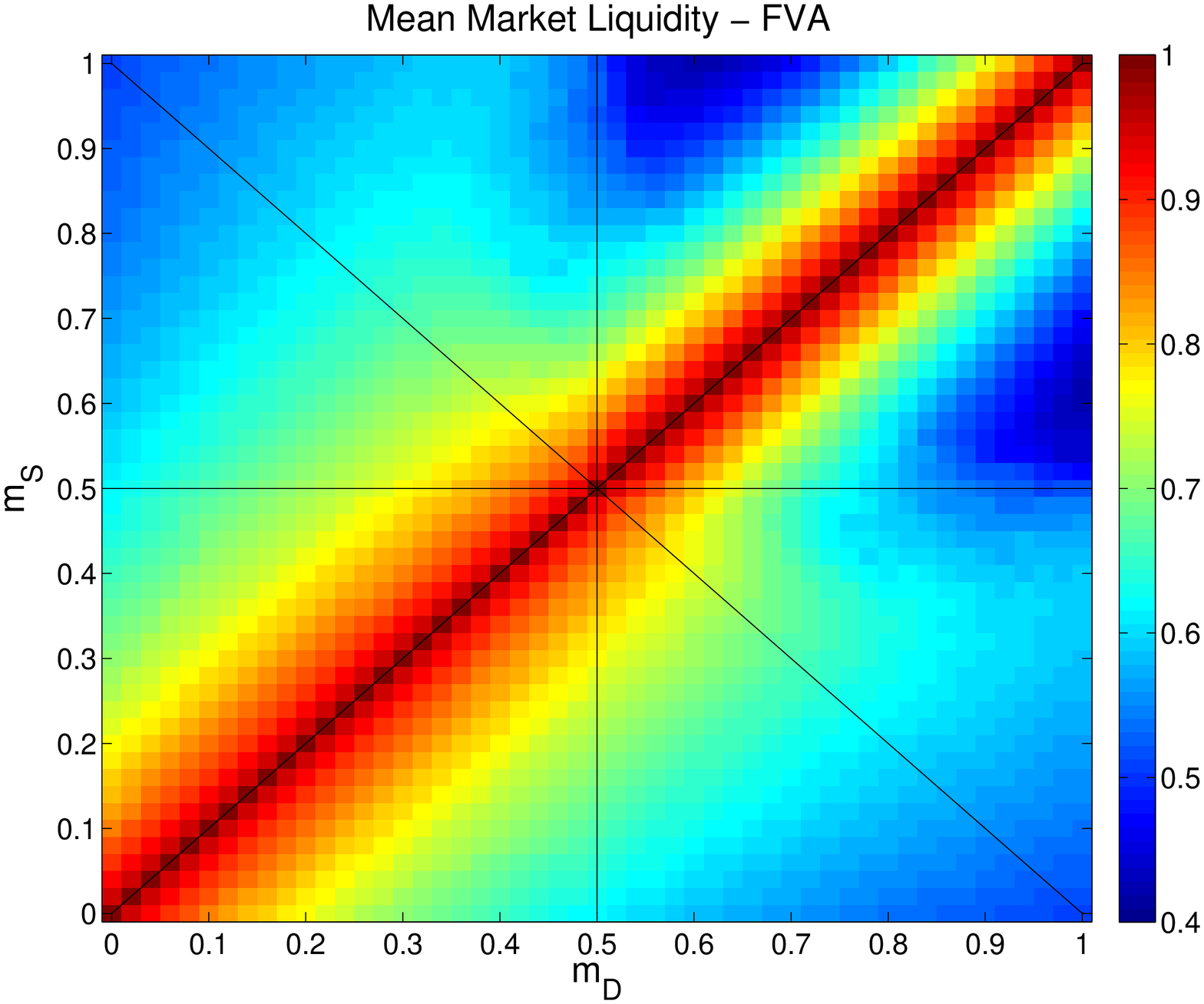}
\includegraphics[width=0.49\textwidth]{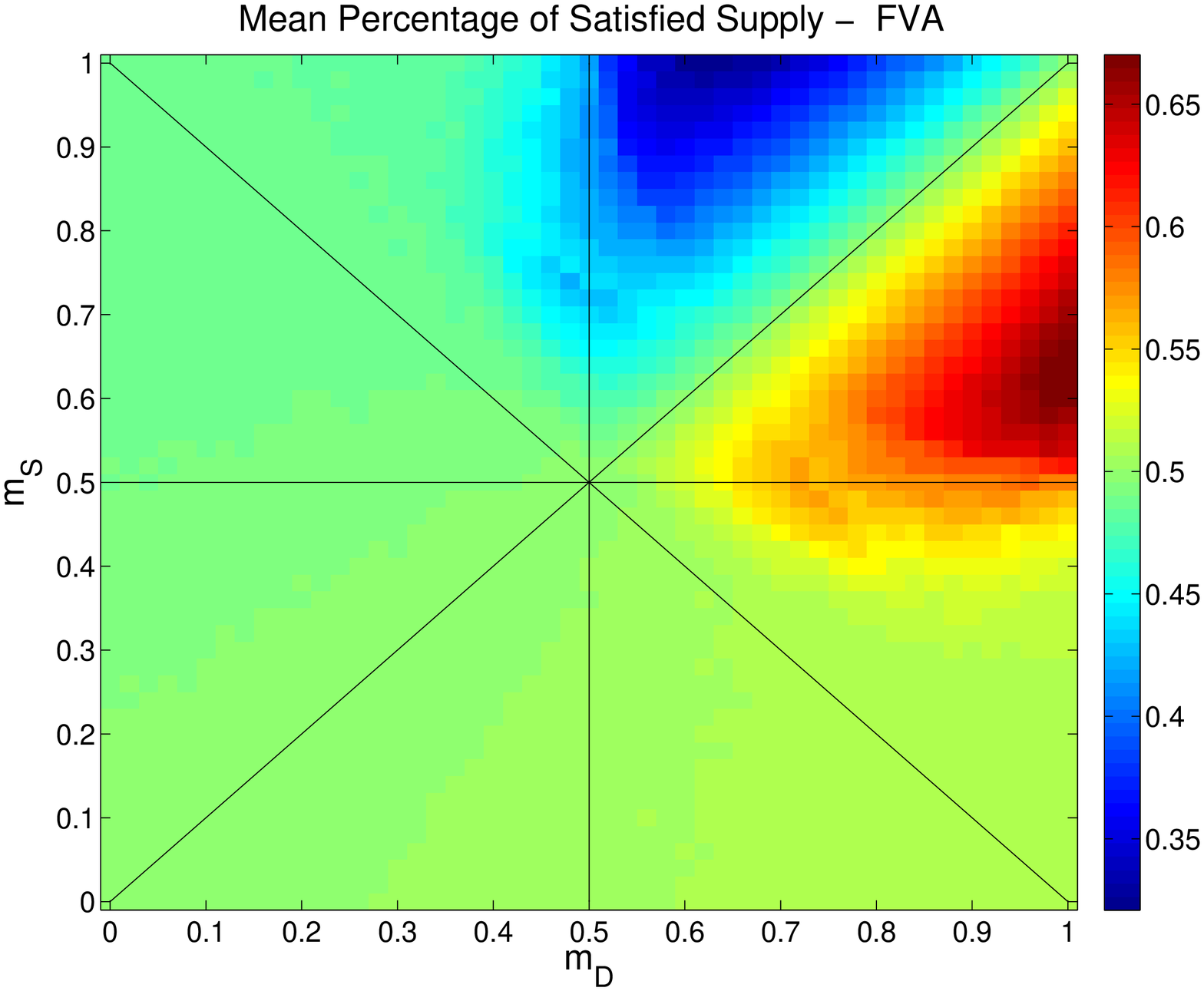}
\caption{Market liquidity expressed as the ratio between investors that participate to the clearing process (clearing area) and those potentially interested in participating to the market (market area). Dispersion of results in space of market confidence $(m_S; m_D)$. Each data point represents the average result of 1000 simulations over 500 periods $t$. Market ratio is represented for for HRT (Upper Left Panel), TRA-S (Upper right Panel) and FVA (Lower Left Panel). Lower right panel: Percentage of satisfied supply for the FVA common knowledge regime.}
\label{Market_ratio_img}
\end{figure}

In particular, our aggregative market matching protocol denotes two areas: one larger area comprises all investors potentially willing to trade (market area); another - smaller -  area comprises all investors that may actually contribute to the clearing process given the bids on the opposite market side (clearing area). The study of the ratio between these two areas (by construction comprised between zero and one) provides some understanding of the timely satisfaction of investors’ orders at each market period $t$, pointing to market liquidity as measured by posted orders satisfaction. While the numerical values are quite in line for all accounting regimes, their organization over the parameter space of market confidence (Figure \ref{Market_ratio_img}) results to be distorted  in the case of FVA, especially around the second quadrant denoting market overconfidence shared by both supply and demand.

Under TRA-S and HRT, market liquidity is symmetric along the diagonal. Orders satisfaction is then increased as long as market confidence is similar on both sides of the market, and maximal where $m_D=m_S$ independently from how conservative or speculative market sentiments are. Orders satisfaction is minimal at the corner where market sentiments on the two sides of the market are opposite, that is, where $(m_D \rightarrow 0) \cap (m_S \rightarrow 1)$ and vice-versa. 

In case of conservative weight attributed to the market price trend, the market matching process works under FVA as well, enabling a high proportion of orders satisfaction for investors willing to exchange. However, under FVA, this result does not hold for the region where at least one market side sentiment tends to be speculative, implying $(m_D>0.5) \cup (m_S>0.5)$ and especially in the area where $(m_D >0.5) \cap (m_S>0.5)$. Here, high market liquidity is maintained only around the diagonal (for market moods by both market sides that are very similar to each other), while the rest of the region shows lower efficiency relative to HRT. This phenomenon results from market instability in that region: both sides are highly confident in the market signal but they do not agree on the precise weight to assign it, thus generating lower levels of market liquidity.
This interpretation is confirmed by studying the percentage of satisfied supply in that region (Figure \ref{Market_ratio_img}). In the part above the diagonal, supply is more overconfident about the market signal than demand, leading to higher levels of unsatisfied supply orders. At the opposite, in the part below the diagonal, demand is more overconfident than supply, leading to lower levels of unsatisfied supply orders. Potential suppliers (buyers) are here overconfident about the market trend relative to potential buyers, willing to selling (buying) at a higher degree than potential buyers (sellers). This generates over supply (demand) that cannot be absorbed by an aggregate matching process based upon disagreement. Orders remain then unsatisfied at a higher degree because of such over-disagreement.
In the other regions, however, the market matching process remains satisfyingly around its expected level of one half (implied by the assumption of uniform distribution on both sides of the market).

\subsection{Information quality}
\label{quality} 
In the context of accounting and economics literature, the linear correlation between market price and market signal (denoted here by $cov[p_t;S_t]$) is generally employed to assess the quality of accounting information (\citealt{lintner1956distribution, demsetz1997economics, lev1999boundaries}).\footnote{The relationship between this hypothesis and an equilibrium approach is acknowledged but not investigated here.} This section provides results for the distributional characteristics of cross-sectional and lagged correlations between these two variables in Tables \ref{cov_tont} and \ref{cov_tmeno1ont}.

\begin{table}[h]
\small
\center
\begin{tabular}{| l | c | c | c | c | c | c |}
\hline
Signal Type & Mean & Min & Q1 & Median & Q3 & Max \\
\hline
HRT (mean) & 0.8177 &0.3988& 0.7330& 0.8191& 0.9059& 0.9895\\
HRT(median) && 0.5030& 0.8261& 0.8891& 0.9444& 0.9919\\
\hline
FVA (mean)& 0.9585 &0.8157& 0.9210& 0.9759& 0.9980& 1\\
FVA (median)&&0.9380& 0.9724& 0.9907& 1& 1\\
\hline
TRA-S (mean) & 0.4803 &-0.0644& 0.2735& 0.4592& 0.6917& 0.9698\\
TRA-S (median) &&-0.0847& 0.3548 & 0.5667 & 0.7828 & 0.9771\\
\hline
 \end{tabular}
\caption{Cross-sectional covariance between cumulated fundamental signal $S_{t}$ and market price $p_t$. Comparison between the distributional characteristics (Mean, Minimum value, First Quartile, Median value, Third Quartile and Maximum value) under the different common knowledge regimes. For each regime, the first line indicates the distributional characteristics of the mean value while the second line indicates the distributional characteristics of the median value.}
\label{cov_tont}
\end{table}

As one can observe, FVA shows high degrees of cross-sectional correlation between market price and fundamental signal, both in absolute terms and compared to the HRT and TRA-S regimes. Some might be then tempted to conclude that FVA provides better accounting information quality. However, as we have seen in the previous sections, this comes at the cost of higher instability, exuberance and vagary imposed to the financial system dynamics. By taking the theoretical accounting regime as benchmark, only the latter shows some negative cross-sectional correlation that is expected when the market price series overshoots-and-reverts around the fundamental signal series. Contrary to the common interpretation of this variable, higher (average) correlation seems then to imply worse systemic performances, with respect to theoretical common knowledge regime benchmark. On this basis, historical cost accounting regime HRT seems to perform better, since its average (median) correlation is less than fair value accounting regime.

The correlation between the market price at time t ($p_t$) and the lagged fundamental signal ($S_{t-1}$) at time $t-1$ (Table \ref{cov_tmeno1ont}) points to the forecast power of the common knowledge information over the market price series one-step-ahead. It is notable that this correlation is unstable over time and contexts, confirming that no trading strategies can reasonably exploit it systematically across situations. Furthermore, the theoretical regime TRA-S shows inferior linear forecast power, surely because of market price overshoot-and-revert effect around its benchmark level. From this perspective, the superior forecast power showed by fair value accounting FVA may be assessed negatively, since it depends on its autocorrelation more than its capacity to drive investment behaviors to fit with benchmark levels of fundamental performance.
\begin{table}[h]
\small
\center
\begin{tabular}{| l | c | c | c | c | c | c |}
\hline
Signal Type & Mean & Min & Q1 & Median & Q3 & Max \\
\hline
HRT (mean)&0.8265 &0.40404& 0.74102& 0.82773& 0.91566& 1\\
HRT(median)&&0.50969& 0.83423& 0.89812& 0.95387& 1\\
\hline
FVA (mean)&0.95928 &0.8167& 0.92969& 0.97717& 0.99853& 1\\
FVA (median)& &0.9387& 0.97309& 0.99184& 1& 1\\
\hline
TRA-S (mean) & 0.51089 &-0.006575& 0.30085& 0.48942& 0.72541& 1\\
TRA-S (median)&&-0.016519& 0.3866& 0.60198& 0.81851& 1\\
\hline
\end{tabular}
\caption{One-period lagged covariance between cumulated fundamental signal $S_{t-1}$ and market price $p_t$. Comparison between the distributional characteristics (Mean, Minimum value, First Quartile, Median value, Third Quartile and Maximum value) under the different common knowledge regimes. For each regime, the first line indicates the distributional characteristics of the mean value while the second line indicates the distributional characteristics of the median value.}
\label{cov_tmeno1ont}
\end{table}

\section{Conclusive Remarks}

Through numerical simulations and visualizations, this paper develops a comprehensive economic analysis of the theoretical model developed by \cite{biondi2012formation}. It assesses market price formation and behavior under alternative common knowledge regimes coupled with various combinations of speculative and fundamentalist beliefs on supply and demand sides, with a view to financial market stability, volatility, exuberance, vagary and liquidity. These systemic properties prove to be sensitive to regulatory regimes for fundamental information provision that correspond to stylized accounting models of reference for financial reporting and prudential regulation.

Through mere observation of linear correlation between market prices and fundamental signals, some might conclude that common knowledge provision regimes based upon mark-to-market measurement of traded security (FVA) represent a better source of public information about the fundamental performance of security-issuing corporate group. However, we show that this high linear correlation comes together with relevant reduction in systemic properties of the financial system as a whole. In particular, FVA-based regimes prove to involve market misplacing and higher volatility. They also imply higher degree of exuberance and errancy (pointing to the disconnection of the market price series from the underlying fundamental signal series of traded security) coupled with lower levels of orders’ satisfaction for both sides of the financial trading (which point to possible liquidity issues). Under FVA-based regimes, therefore, there exists higher likelihood and materiality of financial market instability and bubbling, related to market price erratic behavior over time and contexts. 

Our results are particularly significant since the systemic properties of common knowledge regimes under investigation do radically depend on investors' relative confidence on the market price dynamics. As the latter factor is not generally under control by regulatory bodies which can shape the common knowledge provision regimes, this dependence implies that adoption and implementation of mark-to-market regimes may increase likelihood of market bubbling and inefficiency when market conditions become overconfident. Our study thus recommends to test the economic consequences of regulatory designs and control systems under a large set of situations through simulation, experiment and exploratory field studies.

\newpage
\appendix
\section*{Appendix 1 - Summary of common knowledge regimes}
Together with the HRT, FVA and TRA-S common knowledge regimes (introduced in Equations \ref{HRT}, \ref{FVA} and \ref{TRAS}) we analyze here two further regimes as robustness check of the results for HRT and TRA-S respectively.

The \textbf{Historical Cost Accounting (HCA)} regime implies an evolving exogenous signal of fundamental performance that is orthogonal to market price dynamics and results from stochastic positive and negative flows (\citealt{Biondi2011PureLogic}):

\begin{equation}
F_t=N[-1;+1] + \epsilon_t \,\,\,\, \forall t	
\end{equation}

The \textbf{Fixed Target Reverting Accounting (TRA-F)} regime implies instead a reverting fundamental performance signal that targets a fixed core value of reference:

\begin{equation}
F_t = - (p_t - F_{t-1}) + \epsilon_t \,\, \forall t
\end{equation}

Again, for these two accounting regimes the same stochastic error (introduced in Equation \ref{epsilont}) is added to account for estimation error, measurement error and other random effects.

Overall, we propose in this paper results for five different mechanisms (summarized in Table \ref{typesofknowledge}) generating common knowledge information inspired from distinctive accounting regimes: two belonging to the historical cost accounting model family; one belonging to the fair value (current value, mark-to-market) accounting model family and two providing a theoretical benchmark derived from equilibrium approaches, involving target-based signaling provision. We refer to the first three regimes as \textit{actual} regimes as they denote stylized practical modes of accounting, while referring to the last two regimes as \textit{theoretical} regimes that are and can be only thought in a vacuum.

\begin{table}[h]
\small
\begin{tabular}{| l | l |}
\hline
Actual accounting regimes & \begin{tabular}{ l | l } Historical Cost & Historical Accounting (HCA)\\ Accounting Family& \\ \cline{2-2} & Historical Random Trend \\ & Accounting (HRT)\\ \hline Current Value (Mark-to-Market) & Fair value accounting (FVA) \\
Accounting family & \\ \end{tabular}\\ 
\hline
Theoretical accounting regimes & Fixed target reverting accounting (TRA-F) \\
\hline
& Stochastic target reverting accounting (TRA-S) \\
\hline
\end{tabular}
\caption{Taxonomy of the common knowledge regimes discussed in this paper.}
\label{typesofknowledge}
\end{table}

\appendix
\section*{Appendix 2 - Data tables (comparisons across common knowledge regimes)}
In this appendix we reproduce the tables displayed in the main text including also the omitted results for the theoretical \textit{Fixed Target Reverting Accounting (TRA-F)} regime and for the actual \textit{Historical Cost Accounting (HCA)} regime. Tables \ref{volatility_witdth_appendix} and  \ref{market_ratio_appendix} are instead only reported here to confirm results observed through other variables in the main text.

\subsection*{Market Pricing}
\begin{table}[h]
\small
\center
\begin{tabular}{| l | c | c | c | c | c | c |}
\hline
Signal Type & Mean $\pm$ Std & Min & Q1 & Median & Q3 & Max\\
\hline
HCA(mean) &1000.251 $\pm$ 3.5212 & 996.986 & 1000.098 & 1000.238 & 1000.377 & 1022.292\\
HCA (median) & & 996.7165 & 1000.098 & 1000.239 & 1000.379 & 1003.872\\
\hline
HRT (mean) & 1000.253 $\pm$ 3.6688 & 996.472 & 1000.096 & 1000.237 & 1000.378 & 1035.581\\
HRT (median) & & 996.736 & 1000.098 & 1000.239 & 1000.383 &1016.071\\
\hline
FVA (mean) & 2.4e+60 $\pm$ 2.1e+61 & 999.557 & 1000.490 & 1000.820 & 2.1e+17 & 5.7e+63\\
FVA (median) & & 997.108 & 1000.483 & 1000.816 & 4.2e+08 & 1.2e+32\\
\hline
TRA-F (mean) & 1000.154 $\pm$ 0.1983 & 999.963 & 1000.071 & 1000.144 & 1000.232 & 1000.394 \\
TRA-F (median) & & 999.962 & 1000.072 & 1000.144 &1000.232 & 1000.398\\
\hline
TRA-S (mean) & 1000.155 $\pm$ 1.7446 & 999.589 & 1000.064 & 1000.154 & 1000.246 & 1000.550 \\
TRA-S (median) & & 999.568 &1000.062 & 1000.154 & 1000.248 &1000.563 \\ 
\hline
\end{tabular}
\caption{Market Prices. Comparison between the distributional characteristics (Mean, Standard Deviation, Minimum value, First Quartile, Median value, Third Quartile and Maximum value) under the different common knowledge regimes. For each regime, the first line indicates the distributional characteristics of the mean value while the second line indicates the distributional characteristics of the median value.}
\label{1Meanmarketprice_appendix}
\end{table}

\newpage
\subsection*{Market Volatility}
\begin{table}[h]
\small
\center
\begin{tabular}{| l | c | c | c | c | c | c |}
\hline
Signal Type & Mean & Min & Q1 & Median & Q3 & Max \\
\hline
HCA(mean) &0.0035 &0.0017 & 0.0021 & 0.0025 & 0.0036 & 0.3205 \\
HCA ($75\%$ Peak) & 0.0043 & 0.0020 & 0.0025 & 0.0031 & 0.0043 & 0.4823 \\
\hline
HRT(mean) &0.0037 &0.0018 & 0.0022 & 0.0027 & 0.0037 & 0.3363 \\
HRT ($75\%$ Peak)& 0.0045 & 0.0021 & 0.0026 & 0.0032 & 0.0045 & 0.4791 \\
\hline
FVA (mean) &1.6436 &0.0003 & 0.0006 & 0.0010 & 2.9661 & 8.7466 \\
FVA ($75\%$ Peak) & 1.7882 & 0.0004 & 0.0007 & 0.0010 & 4.3298 & 8.7466 \\
\hline
TRA-F(mean) &0.0002 &0.0001 & 0.0002 & 0.0002 & 0.0002 & 0.0003 \\
TRA-F ($75\%$ Peak) & 0.0002 & 0.0001& 0.0002& 0.0002 & 0.0003& 0.0004 \\
\hline
TRA-S (mean) &0.0017 &0.0012 & 0.0014 & 0.0016 & 0.0019 & 0.0035 \\
TRA-S ($75\%$ Peak) & 0.0021 & 0.0014 & 0.0017 & 0.0020 & 0.0024 & 0.0042 \\
\hline
\end{tabular}
\caption{First Lines: Mean volatility $v_t$ for market price series. Second Lines: Volatility likelihood at $75\%$ of the peak point. Comparison between the distributional characteristics (Mean, Minimum value, First Quartile, Median value, Third Quartile and Maximum value) under the different common knowledge regimes.}
\label{volatility_price_appendix}
\end{table}

The volatility width $W_v$ around its median ($Q2_v$) further comforts the results that regard the Market volatility. Lets define:
\begin{equation}
W_v=\frac{Q3_v-Q1_v}{Q2_v} \forall t
\label{vol_width}
\end{equation}
where Q1, Q2 and Q3 respectively represents the first quartile, the median value and the third quartile of the market volatility distribution.
We can observe in Table \ref{volatility_witdth_appendix} that historical cost regimes remain in line with performance by theoretical regimes while the fair value regime shows anomalously high values of volatility width.

\begin{table}[h]
\small
\center
\begin{tabular}{| l | c | c | c | c | c | c |}
\hline
Signal Type & Mean & Min & Q1 & Median & Q3 & Max \\
\hline
HCA &0.5658 &0.4821& 0.5495& 0.5647& 0.5809& 1.2735\\
\hline
HRT &0.5668 &0.4795& 0.5502& 0.5655& 0.5818& 1.0926\\
\hline
FVA& 28.1821 &0 & 0.5529& 0.6397& 0.6976& 3809.6\\
\hline
TRA-F&0.6081 &0.4867& 0.5667& 0.6019& 0.6498& 0.7624\\
\hline
TRA-S&0.5513 &0.4870& 0.5362& 0.5507& 0.5658& 0.6376\\
\hline
 \end{tabular}
\caption{Volatility width $W_v$ of the signal around its median (see Equation \ref{vol_width} for definition). Comparison between the distributional characteristics (Mean, Minimum value, First Quartile, Median value, Third Quartile and Maximum value) over the different common knowledge regimes (mean behaviour).}
\label{volatility_witdth_appendix}
\end{table}

\newpage

\subsection*{Market Exuberance and Vagary}

\begin{table}[h]
\small
\center
\begin{tabular}{| l | c | c | c | c | c | c |}
\hline
Signal Type & Mean $\pm$ Std & Min & Q1 & Median & Q3 & Max \\
\hline
HCA (75\% Peak)& 0.0022 $\pm$ 0.0041 &0.0002 & 0.0014 & 0.0018& 0.0022& 0.2647\\
HCA ($\overline{exub}$) & 2.00e-05 &2.52e-06& 1.35e-05& 1.62e-05& 1.89e-05& 0.0018\\
\hline
HRT (75\% Peak)& 0.0023 $\pm$ 0.0043 &0.0002 & 0.0015& 0.0019 & 0.0023& 0.2922\\
HRT ($\overline{exub}$) & 2.10e-05 &2.84e-06& 1.41e-05& 1.69e-05& 1.98e-05& 0.0019\\
\hline
FVA (75\% Peak)& 0.1018 $\pm$ 0.0160 &9.68e-05& 0.0003& 0.0003& 0.1085& 0.8660\\
FVA ($\overline{exub}$) & 2.22e-04 &9.36e-07& 2.65e-06& 2.72e-06& 2.81e-04& 1.73e-03\\
\hline
TRA-F(75\% Peak) &0.0005 $\pm$ 0.0003 &2.49e-05& 0.0003& 0.0005 & 0.0007 & 0.0012\\
TRA-F($\overline{exub}$) & 2.09e-06 &2.83e-07& 1.41e-06& 2.12e-06& 2.78e-06& 3.67e-06\\
\hline
TRA-S (75\% Peak) & 0.0019 $\pm$ 0.0033 &0.0002 & 0.0013& 0.0019& 0.0025& 0.0035\\
TRA-S($\overline{exub}$) &1.69e-05 & 2.79e-06& 1.24e-05& 1.72e-05& 2.12e-05& 2.98e-05\\
\hline
\end{tabular} 
\caption{First line: Expected maximum distance $d_t$ at 75\% likelihood (see Equation \ref{Q3dt} for definition). Second Line: Mean exuberance (see Equation \ref{mean_exuberance} for definition). Comparison between the distributional characteristics (Mean, Standard Deviation, Minimum value, First Quartile, Median value, Third Quartile and Maximum value) under the different common knowledge regimes.}
\label{exp_max_dist_appendix}
\end{table}

\begin{table}[h]
\small
\center
\begin{tabular}{| l | c | c | c | c | c | c |}
\hline
\hline
Signal Type & Mean & Min & Q1 & Median & Q3 & Max \\
\hline
HCA ($\%$) &0.2986 &0& 0.2562& 0.3154& 0.3575& 0.5108\\ 
HCA (length) &16 &0& 15& 16& 17& 22\\
\hline
HRT($\%$) &0.3022 &0& 0.2605& 0.3186& 0.3602& 0.5078\\
HRT(length) &16 &0& 16 & 16 & 17 & 21\\
\hline
FVA($\%$) &0.2722 &0& 0.0456& 0.3608& 0.3992& 0.5180\\ 
FVA(length) &14 &0& 16 & 17& 18 & 36\\
\hline
TRA-F($\%$) &0.2618 &0& 0.1888& 0.2892& 0.3436& 0.4822\\
TRA-F(length) &16 &0& 16 & 17& 17& 20\\
\hline
TRA-S($\%$) &0.24186 &0& 0.166& 0.2606& 0.3246& 0.4446\\
TRA-S(length) &16 &0& 15& 16& 16& 22\\
\hline
 \end{tabular}
\caption{First Lines: Percentage of time in which the market price evolution is dissociated from the cumulated fundamental signal $S_t$. Second Lines: Mean Length of the dissociation periods. Comparison between the distributional characteristics (Mean, Minimum value, First Quartile, Median value, Third Quartile and Maximum value) under the different common knowledge regimes.}
\label{dissperc_appendix}
\end{table}

\newpage

\subsection*{Market Liquidity}
\begin{table}[h]
\small
\center
\begin{tabular}{| l | c | c | c | c | c | c |}
\hline
Signal Type & Mean $\pm$ Std & Min & Q1 & Median & Q3 & Max \\
\hline
HCA (mean) & 0.7234 $\pm$ 0.2173 &0.4916& 0.6261& 0.7121& 0.8153& 0.9944\\
HCA (median) &&0.4985& 0.6728& 0.7767& 0.8822& 1\\
\hline
HRT (mean)&0.7232 $\pm$ 0.2166 &0.4899 & 0.6263 & 0.7124 & 0.8149 & 0.9944\\
HRT (median) &&0.4965 & 0.6728 & 0.7767 & 0.8823 & 1\\
\hline
FVA (mean)&0.6925 $\pm$ 0.1932 &0.4205 & 0.5918 & 0.6648 & 0.7854& 0.9923\\
FVA (median) &&0.3727& 0.6174& 0.7052& 0.8372& 1\\
\hline
TRA-F (mean)&0.7205 $\pm$ 0.2202 &0.4855& 0.6188& 0.7101& 0.8153& 0.9921\\
TRA-F (median) &&0.4849& 0.6627& 0.7751& 0.8828 & 1\\
\hline
TRA-S (mean)&0.7208 $\pm$ 0.2179 &0.4792& 0.6184& 0.7111& 0.8169& 0.9944\\
TRA-S(median) &&0.4831& 0.6638& 0.7761& 0.8838& 1\\
\hline
 \end{tabular}
\caption{Ratio between the agents that participate to the clearing process and the total of those willing to trade. Comparison between the distributional characteristics (Mean, Standard Deviation, Minimum value, First Quartile, Median value, Third Quartile and Maximum value) under the different common knowledge regimes. For each regime, the first line indicates the distributional characteristics of the mean value while the second line indicates the distributional characteristics of the median value.}
\label{market_ratio_appendix}
\end{table}

\newpage

\subsection*{Information Quality}
\begin{table}[h]
\small
\center
\begin{tabular}{| l | c | c | c | c | c | c |}
\hline
Signal Type & Mean & Min & Q1 & Median & Q3 & Max \\
\hline
HCA (mean) & 0.8182 &0.4124& 0.7335& 0.8186& 0.9067& 0.9897\\
HCA (median) && 0.5306 & 0.8269 & 0.8881 & 0.9453& 0.9922\\
\hline
HRT (mean) & 0.8177 &0.3988 & 0.7330& 0.8191& 0.9059& 0.9895\\
HRT(median) && 0.5029& 0.8261& 0.8891& 0.9444& 0.9919\\
\hline
FVA (mean)& 0.9585 &0.8157& 0.9290& 0.9759& 0.9981& 1\\
FVA (median)&&0.9380& 0.9724 & 0.9907 & 1& 1\\
\hline
TRA-F (mean) & 0.2845 &-0.3577& -0.0258 & 0.2391& 0.5856 & 0.9695\\
TRA-F (median)& &-0.5335 & -0.0492 & 0.3230 & 0.7066& 0.9756\\
\hline
TRA-S (mean) & 0.4803 &-0.0645& 0.2735& 0.4592& 0.6917& 0.9697\\
TRA-S (median) &&-0.0847 & 0.3548 & 0.5667& 0.7828& 0.9771\\
\hline
 \end{tabular}
\caption{Cross-sectional covariance between cumulated fundamental signal $S_{t}$ and market price $p_t$. Comparison between the distributional characteristics (Mean, Minimum value, First Quartile, Median value, Third Quartile and Maximum value) under the different common knowledge regimes. For each regime, the first line indicates the distributional characteristics of the mean value while the second line indicates the distributional characteristics of the median value.}
\label{cov_tont_appendix}
\end{table}
\begin{table}[h]
\small
\center
\begin{tabular}{| l | c | c | c | c | c | c |}
\hline
Signal Type & Mean & Min & Q1 & Median & Q3 & Max \\
\hline
HCA (mean)&0.8270 &0.4177 & 0.7414 & 0.8277 & 0.9165 & 1\\
HCA (median)&&0.5339& 0.83508& 0.8972 & 0.9542 & 1\\
\hline
HRT (mean)&0.8265 &0.4040 & 0.7410 & 0.8277& 0.9156& 1\\
HRT(median)&&0.5097& 0.8342& 0.8981& 0.9539 & 1\\
\hline
FVA (mean)&0.9593 &0.8167& 0.9297& 0.9772& 0.9985& 1\\
FVA (median)& &0.9387& 0.9731& 0.9918& 1& 1\\
\hline
TRA-F (mean) &0.3117 &-0.3464 & 0.0002 & 0.2680 & 0.6183 & 1\\
TRA-F (median)&&-0.5225& -0.0193& 0.3598& 0.7451& 1\\
\hline
TRA-S (mean) & 0.5109 &-0.0066 & 0.3008& 0.4894 & 0.7254 & 1\\
TRA-S (median)&&-0.0165& 0.3866 & 0.6020 & 0.8185 & 1\\
\hline
\end{tabular}
\caption{One-period lagged covariance between cumulated fundamental signal $S_{t-1}$ and market price $p_t$. Comparison between the distributional characteristics (Mean, Minimum value, First Quartile, Median value, Third Quartile and Maximum value) under the different common knowledge regimes. For each regime, the first line indicates the distributional characteristics of the mean value while the second line indicates the distributional characteristics of the median value.}
\label{cov_tmeno1ont_appendix}
\end{table}
\newpage

\cleardoublepage
\addcontentsline{toc}{section}{\refname}
\bibliographystyle{apalike}
\bibliography{BiondiRighi}

\end{document}